\documentclass[fleqn,usenatbib]{mnras}

\usepackage[T1]{fontenc}

\DeclareRobustCommand{\VAN}[3]{#2}
\let\VANthebibliography\thebibliography
\def\thebibliography{\DeclareRobustCommand{\VAN}[3]{##3}\VANthebibliography}

\usepackage{graphicx}
\usepackage{amsmath}
\usepackage{amssymb}
\usepackage{upgreek}
\usepackage{anyfontsize}
\usepackage{pifont}
\usepackage{placeins}
\usepackage{orcidlink}

\usepackage{lineno}

\newcommand{\spectralline}[3]{\ion{#1}{#2}~$\uplambda#3$~\AA}
\newcommand{\fspectralline}[3]{[\ion{#1}{#2}]~$\uplambda#3$~\AA}
\newcommand{\shortspecline}[2]{\ion{#1}{#2}}
\newcommand{\shortfspecline}[2]{[\ion{#1}{#2}]}
\newcommand{\Ha}{H$\upalpha$}

\newcommand{\msun}{\mathrm{M}_\odot}
\newcommand{\mathdash}{~\mbox{--}~}

\newcommand{\citeme}{\citetalias{Callow2024a}}

\newcommand{\mycomment}[1]{}
\newcommand{\change}[1]{#1}
\newcommand{\desichange}[1]{#1}

\newcommand{\mnraschange}[1]{#1}

\newcommand{\sdssgalrate}{$R_\mathrm{G}=3.6~^{+2.6}_{-1.8}~(\mathrm{statistical})~^{+5.1}_{-0.0}~(\mathrm{systematic})\times10^{-6}~\mathrm{galaxy}^{-1}~\mathrm{yr}^{-1}$}

\newcommand{\sdssvolrate}{$R_\mathrm{V}=7~^{+20}_{-5}~(\mathrm{statistical})~^{+10}_{-0.0}~(\mathrm{systematic})\times10^{-9}~\mathrm{Mpc}^{-3}~\mathrm{yr}^{-1}$}

\newcommand{\lowzgalrate}{$R_\mathrm{G}=1.6~^{+3.8}_{-1.4}\times10^{-6}~\mathrm{galaxy}^{-1}~\mathrm{yr}^{-1}$}
\newcommand{\lowzmassrate}{$R_\mathrm{M}=7~^{+16}_{-6}\times10^{-18}~\mathrm{M_\odot^{-1}}~\mathrm{yr}^{-1}$}
\newcommand{\lowzvolrate}{$R_\mathrm{V}=1.8~^{+4.5}_{-1.5}\times10^{-9}~\mathrm{Mpc}^{-3}~\mathrm{yr}^{-1}$}

\newcommand{\figref}[1]{Fig. \ref{#1}}
\newcommand{\tabref}[1]{Table \ref{#1}}
\newcommand{\secref}[1]{Section \ref{#1}}
\newcommand{\equref}[1]{equation (\ref{#1})}

\title[ECLEs in BOSS LOWZ]{The rate of extreme coronal line emitters in the Baryon Oscillation Spectroscopic Survey LOWZ sample}

\author[J. Callow et al.]{
\parbox{\textwidth}{
J.~Callow$^{\orcidlink{0000-0002-0804-9533}}$,$^{1}$\thanks{E-mail: joe.callow@port.ac.uk}
O.~Graur$^{\orcidlink{0000-0002-4391-6137}}$,$^{1,2}$
P.~Clark$^{\orcidlink{0000-0002-6576-7400}}$,$^{3,1}$
A.~G.~Kim$^{\orcidlink{0000-0001-6315-8743}}$,$^{4}$
B.~O'Connor$^{\orcidlink{0000-0002-9700-0036}}$,$^{5}$
J.~Aguilar$^{\orcidlink{0000-0003-0822-452X}}$,$^{6}$
S.~Ahlen$^{\orcidlink{0000-0001-6098-7247}}$,$^{7}$
D.~Bianchi$^{\orcidlink{0000-0001-9712-0006}}$,$^{7}$
D.~Brooks$^{\orcidlink{0000-0002-8458-5047}}$,$^{8}$
A.~de la Macorra$^{\orcidlink{0000-0002-1769-1640}}$,$^{9}$
A.~Dey$^{\orcidlink{0000-0002-4928-4003}}$,$^{10}$
P.~Doel$^{\orcidlink{0000-0002-6397-4457}}$,$^{8}$
J.~E.~Forero-Romero$^{\orcidlink{0000-0002-2890-3725}}$,$^{11,12}$
E.~Gaztañaga$^{\orcidlink{0000-0001-9632-0815}}$,$^{13,1,14}$
S.~Gontcho A Gontcho$^{\orcidlink{0000-0003-3142-233X}}$,$^{4}$
G.~Gutierrez$^{\orcidlink{0000-0003-0825-0517}}$,$^{15}$
R.~Kehoe,$^{16}$
A.~Lambert$^{\orcidlink{0009-0003-5658-2601}}$,$^{4}$
M.~Landriau$^{\orcidlink{0000-0003-1838-8528}}$,$^{4}$
L.~Le~Guillou$^{\orcidlink{0000-0001-7178-8868}}$,$^{17}$
A.~Meisner$^{\orcidlink{0000-0002-1125-7384}}$,$^{10}$
R.~Miquel$^{\orcidlink{0000-0002-6610-4836}}$,$^{18,19}$
J.~Moustakas$^{\orcidlink{0000-0002-2733-4559}}$,$^{20}$
F.~Prada$^{\orcidlink{0000-0001-7145-8674}}$,$^{21}$
I.~P\'erez-R\`afols$^{\orcidlink{0000-0001-6979-0125}}$,$^{22}$
G.~Rossi,$^{23}$
E.~Sanchez$^{\orcidlink{0000-0002-9646-8198}}$,$^{24}$
M.~Schubnell$^{\orcidlink{0000-0001-9504-2059}}$,$^{25,26}$
H.~Seo$^{\orcidlink{0000-0002-6588-3508}}$,$^{27}$
D.~Sprayberry,$^{10}$
G.~Tarl\'{e}$^{\orcidlink{0000-0003-1704-0781}}$,$^{26}$
B.~A.~Weaver,$^{10}$
and H.~Zou$^{\orcidlink{0000-0002-6684-3997}}$$^{28}$
}
\vspace{0.4cm}
\\
\parbox{\textwidth}{
$^{1}$ Institute of Cosmology \& Gravitation, University of Portsmouth, Dennis Sciama Building, Portsmouth, PO1 3FX, UK\\
$^{2}$ Department of Astrophysics, American Museum of Natural History, New York, NY 10024, USA\\
$^{3}$ School of Physics and Astronomy, University of Southampton, Southampton, SO17 1BJ, UK\\
$^{4}$ Lawrence Berkeley National Laboratory, 1 Cyclotron Road, Berkeley, CA 94720, USA\\
$^{5}$ Department of Physics, Carnegie Mellon University, 5000 Forbes Avenue, Pittsburgh, PA 15213, USA\\
$^{6}$ Physics Dept., Boston University, 590 Commonwealth Avenue, Boston, MA 02215, USA\\
$^{7}$ Dipartimento di Fisica ``Aldo Pontremoli'', Universit\`a degli Studi di Milano, Via Celoria 16, I-20133 Milano, Italy\\
$^{8}$ Department of Physics \& Astronomy, University College London, Gower Street, London, WC1E 6BT, UK\\
$^{9}$ Instituto de F\'{\i}sica, Universidad Nacional Aut\'{o}noma de M\'{e}xico,  Circuito de la Investigaci\'{o}n Cient\'{\i}fica,Ciudad Universitaria, Cd. de M\'{e}xico  C.~P.~04510,  M\'{e}xico\\
$^{10}$ NSF NOIRLab, 950 N. Cherry Ave., Tucson, AZ 85719, USA\\
$^{11}$ Departamento de F\'isica, Universidad de los Andes, Cra. 1 No. 18A-10, Edificio Ip, CP 111711, Bogot\'a, Colombia\\
$^{12}$ Observatorio Astron\'omico, Universidad de los Andes, Cra. 1 No. 18A-10, Edificio H, CP 111711 Bogot\'a, Colombia\\
$^{13}$ Institut d'Estudis Espacials de Catalunya (IEEC), c/ Esteve Terradas 1, Edifici RDIT, Campus PMT-UPC, 08860 Castelldefels, Spain\\
$^{14}$ Institute of Space Sciences, ICE-CSIC, Campus UAB, Carrer de Can Magrans s/n, 08913 Bellaterra, Barcelona, Spain\\
$^{15}$ Fermi National Accelerator Laboratory, PO Box 500, Batavia, IL 60510, USA\\
$^{16}$ Department of Physics, Southern Methodist University, 3215 Daniel Avenue, Dallas, TX 75275, USA\\
$^{17}$ Sorbonne Universit\'{e}, CNRS/IN2P3, Laboratoire de Physique Nucl\'{e}aire et de Hautes Energies (LPNHE), FR-75005 Paris, France\\
$^{18}$ Instituci\'{o} Catalana de Recerca i Estudis Avan\c{c}ats, Passeig de Llu\'{\i}s Companys, 23, 08010 Barcelona, Spain\\
$^{19}$ Institut de F\'{i}sica d’Altes Energies (IFAE), The Barcelona Institute of Science and Technology, Edifici Cn, Campus UAB, 08193, Bellaterra (Barcelona), Spain\\
$^{20}$ Department of Physics and Astronomy, Siena College, 515 Loudon Road, Loudonville, NY 12211, USA\\
$^{21}$ Instituto de Astrof\'{i}sica de Andaluc\'{i}a (CSIC), Glorieta de la Astronom\'{i}a, s/n, E-18008 Granada, Spain\\
$^{22}$ Departament de F\'isica, EEBE, Universitat Polit\`ecnica de Catalunya, c/Eduard Maristany 10, 08930 Barcelona, Spain\\
$^{23}$ Department of Physics and Astronomy, Sejong University, 209 Neungdong-ro, Gwangjin-gu, Seoul 05006, Republic of Korea\\
$^{24}$ CIEMAT, Avenida Complutense 40, E-28040 Madrid, Spain\\
$^{25}$ Department of Physics, University of Michigan, 450 Church Street, Ann Arbor, MI 48109, USA\\
$^{26}$ University of Michigan, 500 S. State Street, Ann Arbor, MI 48109, USA\\
$^{27}$ Department of Physics \& Astronomy, Ohio University, 139 University Terrace, Athens, OH 45701, USA\\
$^{28}$ National Astronomical Observatories, Chinese Academy of Sciences, A20 Datun Rd., Chaoyang District, Beijing, 100012, P.R. China\\
}
}

\date{Accepted XXX. Received YYY; in original form ZZZ}

\pubyear{2024}

\usepackage{newtxtext,newtxmath}

\begin{document}
\label{firstpage}
\pagerange{\pageref{firstpage}--\pageref{lastpage}}
\maketitle

\defcitealias{Callow2024a}{C24}

\begin{abstract}

Extreme coronal line emitters (ECLEs) are a rare class of galaxy that exhibit strong, high-ionization iron coronal emission lines in their spectra.
In some cases, these lines are transient and may be the result of tidal disruption event (TDEs).
To test this connection, we calculate the rate of variable ECLEs (vECLEs) at redshift $\sim0.3$.
We search for ECLEs in the Baryon Oscillation Spectroscopic Survey (BOSS) LOWZ sample and discover two candidate ECLEs.
Using follow-up spectra from the Dark Energy Spectroscopic Instrument and Gemini Multi-Object Spectrograph, and mid-infrared observations from the \textit{Wide-field Infrared Survey Explorer}, we determine that one of these galaxies is a vECLE.
Using this galaxy, we calculate the galaxy-normalized vECLE rate at redshift $\sim0.3$ to be \lowzgalrate and the mass-normalized rate to be \lowzmassrate.
This is then converted to a volumetric rate of \lowzvolrate.
Formally, the LOWZ vECLE rates are $2\mathdash4$ times lower than the rates calculated from the Sloan Digital Sky Survey Legacy sample at redshift $\sim0.1$.
However, given the large uncertainties on both measurements, they are consistent with each other at $1\upsigma$.
Both the galaxy-normalized and volumetric rates are one to two orders of magnitude lower than TDE rates from the literature, consistent with vECLEs being caused by $5\mathdash20$ per cent of all TDEs.

\end{abstract}


\section{Introduction}
\label{sec:intro}

A tidal disruption event (TDE) is a high-energy transient phenomenon, in which a star is gravitationally shredded by a supermassive black hole's (SMBH) tidal forces \citep{Hills1975}.
This causes roughly half of the star's matter to fall onto the black hole and form a temporary accretion disc, during which at least one flare of electromagnetic radiation is produced.
TDEs can produce flares with wavelengths across the electromagnetic spectrum, with examples of TDEs producing X-ray \citep{Saxton2021}, optical/ultraviolet (UV) \citep{vanVelzen2020}, mid-infrared (MIR) \citep{Masterson2024}, and radio emission \citep{Alexander2020}.
Known accretion physics can explain the X-ray and radio emission \citep{Rees1988,Giannios2011}.
However, the optical/UV emission is less well understood.
It may be caused by shocks from the bound stream of disrupted matter colliding with itself as it circularizes around the black hole \citep{Piran2015} or reprocessing of the X-ray emission by material surrounding the black hole \citep{Guillochon2013,Roth2016}.
The MIR emission is created by dust reprocessing the UV radiation \citep{Lu2016,vanVelzen2021b}.

An important aspect of TDEs is the rate at which they occur.
TDE rates derived using standard `loss cone' theory are typically of the order $\sim10^{-4}~\mathrm{galaxy}^{-1}~\mathrm{yr}^{-1}$ \citep{Magorrian1999,Wang2004,Stone2016}, although more recent work on modifying this theory has produced different rates.
\desichange{\citet{Teboul2023} considered interactions that are only significant in the dense stellar environments surrounding SMBHs, such as strong scatterings and collisions between stars.
These interactions can cause stars to be ejected from stellar populations before they are tidally disrupted.
This effect could reduce the theoretical TDE rate by as much as an order of magnitude.}
TDE rates derived from observational studies span a similar range, with values varying from $\sim10^{-5}~\mathrm{galaxy}^{-1}~\mathrm{yr}^{-1}$ \citep{Donley2002,Khabibullin2014,vanVelzen2014,Holoien2016,Yao2023,Masterson2024} to $\sim10^{-4}~\mathrm{galaxy}^{-1}~\mathrm{yr}^{-1}$ \citep{Esquej2008,Maksym2013,Hung2018,vanVelzen2018}.
Both sets of rates come from studies of X-ray and optical/UV surveys, although a rate of $2.0\pm0.3\times10^{-5} ~\mathrm{galaxy}^{-1}~\mathrm{yr}^{-1}$ was also calculated from a MIR survey \citep{Masterson2024}.
It is unclear if the discrepancy between theoretical and observational rates is due to theoretical rate calculations \desichange{being incomplete regarding the modelling of the processes behind tidal disruptions} or whether TDEs are being missed by surveys, which would artificially lower the observational rate.

A possible signature of TDE activity in a dusty environment is a set of strong, high-ionization iron coronal emission lines.
These were first observed in a rare class of galaxies known as extreme coronal line emitters (ECLEs), which were discovered in the Sloan Digital Sky Survey (SDSS) Legacy Survey \citep{Komossa2008,Wang2011}, and were noted as unusual due to these \desichange{coronal lines (CLs)} and broad Balmer features that indicated a strong ionizing continuum.
The CLs present were \fspectralline{Fe}{vii}{3759}, \fspectralline{Fe}{vii}{5160}, \fspectralline{Fe}{vii}{5722}, \fspectralline{Fe}{vii}{6088}, \fspectralline{Fe}{x}{6376}, \fspectralline{Fe}{xi}{7894}, and \fspectralline{Fe}{xiv}{5304} (hereafter, the higher ionization CLs will be referred to as \shortfspecline{Fe}{x}, \shortfspecline{Fe}{xi}, and \shortfspecline{Fe}{xiv}, respectively).
A search through the seventh SDSS data release \citep[DR7; ][]{Abazajian2009} by \citet{Wang2012} discovered a total of seven ECLEs, which showed a range of properties.
Each of the ECLEs showed a different combination of CLs, and there were also differences in the presence and strengths of other emission lines (e.g. \spectralline{He}{ii}{4686} and \fspectralline{O}{iii}{5007}).

\change{Long term observations of ECLEs revealed that they belonged to two subclasses depending on their variability.
Follow-up spectra taken $\sim10$ and $20$ yrs after the SDSS spectra revealed that five of the ECLEs had fading CLs, whereas the CLs remained constant in the other two ECLEs \citep{Wang2012,Yang2013,Clark2024}.
The spectra also showed that the \fspectralline{O}{iii}{5007} emission lines in the variable ECLEs (vECLEs) had increased in strength over the $\sim20$ yrs, whereas they had not changed in the non-vECLEs.
}

It was also noted that the MIR evolution of the vECLEs differed significantly from that of the  other ECLEs \citep{Dou2016,Clark2024,Hinkle2024}.
The MIR emission from the ECLEs in which the CLs did not fade over time remained roughly constant, whereas the vECLEs showed a long-term decline in the MIR.
This difference was particularly noticeable when comparing the $W1-W2$ colour index of the ECLEs, where $W1$ and $W2$ are the bands observed by the \textit{Wide-field Infrared Survey Explorer} (\textit{WISE}).
Again, the non-vECLEs showed little variation and remained above the $W1-W2>0.8$ cut used to discern AGNs from non-AGNs \citep{Stern2012}.
However, the vECLEs all showed a decline from an AGN-like state to a non-AGN-like state, hinting at accretion being the energy source powering the vECLEs.

Due to the high energies required to ionize iron to the levels observed, TDEs, supernovae (SNe), and AGNs were proposed as possible progenitors.
All three of these phenomena have been observed producing long-lasting CLs \citep{Gelbord2009,Smith2009,Fransson2014}, but differences between them can be seen in their duration and luminosities.
\change{CLs produced by SNe are typically weaker than those observed in ECLEs \citep{Komossa2009,Wang2011}, and usually fade much faster than the few years over which the CLs in the \citet{Wang2012} vECLEs persisted \citep{Palaversa2016}.
The Type IIn SN 2005ip developed CLs that were still detectable $\sim3000$ d after they were first observed \citep{Smith2009}, which is more similar to the timescales observed in the vECLEs.
However, these CLs were still much weaker than those observed in ECLEs.}

\change{AGNs also produce CLs that are weaker than observed in ECLEs.
This is most clearly seen when comparing the strength of the CLs to the \fspectralline{O}{iii}{5007} line.
In AGNs, the CLs' luminosities are typically only a few per cent of the \shortfspecline{O}{iii} line \citep{Nagao2000,Gelbord2009,Rose2015,Cerqueira-Campos2020}, whereas in the ECLEs, they have comparable luminosities \citep{Wang2012}.
AGN variation is more erratic and on shorter timescales than observed in the vECLEs and the amplitude of their variation is typically only a few per cent of their total luminosity \citep{Hawkins2002}.}

\change{These differences in CL strengths and the timescales over which the variation occurs make SNe and AGNs less likely to be the progenitors of vECLEs.
TDEs have been predicted to produce sufficient high-energy emission to sustain CLs of this strength for the lengths of time observed in vECLEs \citep{Komossa2008,Yang2013} and are therefore a likely candidate.}

\citet[][hereafter \citeme]{Callow2024a} performed the first full vECLE rate calculation by repeating the search for ECLEs in SDSS Legacy, but using the updated DR17 \citep{Abdurro'uf2022}.
They discovered nine more galaxies with strong CLs.
Follow-up spectra and photometric observations indicated that none of these galaxies had transient CLs.
Using the sample of five vECLEs originally discovered in SDSS and detected once more by \citeme, the authors calculated the galaxy-normalized rate of vECLEs in SDSS Legacy to be \sdssgalrate, which was converted to a volumetric rate of \sdssvolrate.
These rates were one to two orders of magnitude lower than TDE rates from observational studies (but consistent with the lowest TDE rates within the measurements' uncertainties), which suggested vECLEs represent emission from a subset of $10\mathdash40$ per cent of TDEs.

Recently, these same CLs have been observed developing in the spectra of photometrically discovered TDEs (AT~2017gge \citep{Onori2022,Wang2022b}, AT~2018bcb \citep{Neustadt2020}, \mnraschange{AT~2018dyk \citep{Clark2025}, AT~2019aalc \citep{Veres2024}, AT~2019avd \citep{Malyali2021}}, TDE~2019qiz \citep{Nicholl2020,Short2023}, TDE~2020vdq \citep{Somalwar2023}, AT~2021dms \citep{Hinkle2024}, AT~2021qth \citep{Yao2023}, AT~2021acak \citep{Li2023}, TDE~2022fpx \citep{Koljonen2024}, and TDE~2022upj \citep{Newsome2024}).
These TDEs were first detected as optical/UV TDEs, with subsequent detections of X-ray emission typically occurring on the order of hundreds of days after peak (AT~2017gge, AT~2018bcb, AT~2018dyk, AT~2019avd, AT~2021acak, TDE~2022fpx, TDE~2022upj).
However, AT~2019qiz showed X-ray emission around the same time as the optical/UV peak.
In addition, the appearance of the CLs relative to the optical/UV peak is different between TDEs, with CLs being detected within $\sim50$ d of the optical/UV peak in AT~2018bcb, AT~2018dyk, AT~2022fpx, and AT~2022upj.
All the other TDEs showed development of CLs between $170\mathdash500$ d after the peak of the TDE, although in some cases, few spectra were observed, so constraining when the CLs developed is difficult.

Though the link between vECLEs and TDEs is now established, more detections are necessary to build substantial samples of ECLEs and conduct population studies.
Though TDEs have been observed at redshifts ranging as high as $z\sim1.2$ \citep{Andreoni2023}, the maximum redshift that an ECLE has been observed at is 0.14, which is slightly higher than the median redshift of SDSS Legacy.
Additionally, TDE rates have been measured out to $z\sim0.2$ and rates of vECLEs have only been measured out to $z\sim0.1$.
Therefore, in this work, we calculate the rate of vECLEs at a higher redshift than done by \citeme\ by searching for ECLEs in the Baryon Oscillation Spectroscopic Survey \citep[BOSS;][]{Dawson2013} LOWZ sample.
In \secref{sec:data}, we outline the observations used in this work, including the LOWZ sample and follow-up spectroscopic and photometric observations.
We discuss the results of our search in \secref{sec:ecle_search} by examining the properties of the galaxies selected and determining the efficiency of our detection algorithm.
In \secref{sec:ecle_sample}, we describe the cuts made to select the ECLEs from our sample of CL galaxies and the follow-up observations used to determine if they are variable.
We find two candidates ECLEs in the LOWZ sample.
By using follow-up spectroscopic and photometric observations, we determine one to be a vECLE, likely created by a TDE.
In \secref{sec:rate_analysis}, using this new vECLE, we calculate the rate at which vECLEs occur at redshift $\sim0.3$.
We calculate a galaxy-normalized rate of \lowzgalrate and a mass-normalized rate of \lowzmassrate.
We explore the relation between ECLE rate and galactic stellar mass, and convert our mass-normalized rate to a volumetric rate of \lowzvolrate.
Comparing these rates to the ECLE rates from \citeme\ and observational TDE rates, we find that our new measurement is formally lower but statistically consistent with both TDE and vECLE rates at redshift $\sim0.1$.
Finally, we summarize our conclusions in \secref{sec:conclusions}.

Throughout this paper, we assume a Hubble-Lema\^itre constant, $H_0$, of $\mathrm{73\ km\ s^{-1}\ Mpc^{-1}}$ and adopt a standard cosmology with $\Omega_\mathrm{m}=0.27$ and $\Omega_\Lambda=0.73$. \desichange{Quoted uncertainties are at the $1\upsigma$ level, unless stated otherwise.}

\section{Data}
\label{sec:data}

Here, we describe the surveys and instruments used in this work and the data-reduction pipelines used to process the data.

\subsection{BOSS LOWZ}
\label{subsec:boss_lowz}

We performed a search for ECLEs in the SDSS BOSS LOWZ sample \citep{Dawson2013}, which \desichange{contains} spectra of galaxies between $0.15<z<0.43$ \desichange{obtained between 2008 and 2014}.
Following \citet{Graur2013}, \citet{Graur2015}, and \citeme, for each spectrum, the only pixels used were those flagged as `good' (0) or `emission line' (40,000,000) by the BOSS pipeline.
If this cut removed 50 per cent or more of the pixels in a given spectrum, then the spectrum was not used.
This resulted in 1486 spectra being rejected, which left a sample of 341,110 galaxies.
\mnraschange{As we describe in sections \ref{sec:ecle_search} and \ref{subsec:ecle_criteria}, below, we detected two candidate vECLEs in this galaxy sample.}

Every galaxy in the BOSS survey has stellar mass and star formation rate (SFR) estimates derived using the Portsmouth group pipeline \citep{Maraston2013}.
This pipeline fitted all spectra with two templates; a luminous red galaxy model and a star forming model.
The masses and SFRs were then derived using the best fitting model, which was assigned using the colour cut defined by \citet{Maraston2013}.

\subsection{Optical spectroscopy}
\label{subsec:op_spec}

We obtained follow-up optical spectra of one of the ECLEs using the Dark Energy Spectroscopic Instrument (DESI) mounted on the Mayall 4-m telescope \citep{DESIcollab2016a,DESIcollab2016b,DESIcollab2022,DESIcollab2024a,DESIcollab2024b}.
DESI is a robotic, fiber-fed, highly multiplexed spectroscopic surveyor which can obtain simultaneous spectra of almost 5000 objects over a  $\sim3$° field \citep{Silber2023,Miller2024,Poppett2024}, the goal of which is to determine the nature of dark energy through the most precise measurement of the expansion history of the universe ever obtained \citep{Levi2013}.

Using the First Data Release (DR1; DESI Collaboration et al., in prep.), there are science Key Papers presenting the two point clustering measurements and validation \citep{DESIcollab2024II}, baryon acoustic oscillations (BAO) measurements from galaxies and quasars \citep{DESIcollab2024III}, and from the Lya forest \citep{DESIcollab2024IV}, as well as a full-shape study of galaxies and quasars \citep{DESIcollab2024V}. There are Cosmological results from the BAO measurements \citep{DESIcollab2024VI} and the full-shape analysis \citep{DESIcollab2024VII}.

The spectrum has the DESI TARGETID 39628342705525273 and was observed as part of the bright galaxy survey \citep{Hahn2023a} during main survey operations \citep{Schlafly2023} and was processed using the custom DESI spectroscopic pipeline \citep{Guy2023}.
When comparing the BOSS and DESI spectra, it is important to note that the instruments have different sized fibres, with diameters of 2 and 1.5 arcsec, respectively \citep{Smee2013,Abareshi2022}.
Therefore, DESI spectra contain slightly less light from the outer regions of the host galaxies despite being centred on the same location.
This may introduce relative changes in line fluxes and ratios depending on the line-emitting regions covered by the fibres.

We also obtained optical spectra of one of the ECLEs using the Gemini Multi-Object Spectrograph \citep[GMOS;][]{Hook2004} on the 8.1-m Gemini North Telescope (Gemini) on Maunakea, Hawai`i.
These were taken on 2023 December 24 in the long-slit spectroscopy mode with a slit width of 1.0 arcsec, using the B480 and R831 gratings, as part of the Gemini program GN-2023B-Q-321 (PI: P. Clark).
Data were reduced using the \verb|DRAGONS| (Data Reduction for Astronomy from Gemini Observatory North and South) reduction package \citep{Labrie2019}, using the standard recipe for GMOS long-slit reductions.
This includes bias correction, flatfielding, wavelength calibration, and flux calibration.
As we did not have telluric standards for this observation, we used the python package \verb|TelFit| \citep{Gullikson2014} to model and remove the telluric absorption features.

In \tabref{tab:obs_conditions}, we summarize the spectroscopic observing conditions for the ECLE spectra.

\begin{table}
    \centering
    \caption{Summary of the spectroscopic observing conditions of the ECLEs.}
    \begin{tabular}{|c|c|c|c|}
        \hline
        SDSS Short Name & MJD & Seeing (arcsec) & Air mass \\
        \hline
        \multicolumn{4}{c}{\bf{BOSS LOWZ}} \\
		SDSS J0113 & 56189 & 1.5 & 1.2 \\
		SDSS J2218 & 56211 & 1.4 & 1.0 \\
        \hline
        \multicolumn{4}{c}{\bf{DESI}} \\
		SDSS J2218 & 59402 & \desichange{0.9} & \desichange{1.0} \\
        \hline
        \multicolumn{4}{c}{\bf{GMOS}} \\
        SDSS J0113 & 60302 & 1.1 & 1.0 \\
        \hline
    \end{tabular}
    \label{tab:obs_conditions}
\end{table}

\subsection{Mid-infrared photometry}
\label{subsec:ir_phot}

We retrieved MIR photometry from the \textit{WISE} satellite for all the objects in our ECLE sample using the ALLWISE \citep{Wright2010} and NEOWISE Reactivation \citep{Mainzer2011,Mainzer2014} data releases.
We processed the data using a custom \verb|python| script \citep{Clark2024} that removed observations that were marked as an upper limit; were taken when the spacecraft was close to the South Atlantic Anomaly or the sky position of the Moon, or were flagged by the \textit{WISE} pipeline as having a low frame quality or suffering from potential `contamination or confusion.'
\citet{Dou2016} showed that the \citet{Wang2012} vECLEs did not show MIR variability during each observation block.
Therefore, a weighted average was used to produce a single magnitude value per filter for each observation block.

\section{ECLE Search}
\label{sec:ecle_search}

In this section, we describe our CL galaxy selection criteria, compare the galaxies selected to the overall LOWZ sample, and calculate the efficiency of our detection algorithm.
We also compare the galaxy samples to the SDSS Legacy sample used by \citeme\ and describe the differences this makes in the properties of the galaxies that we investigate in this work.

\subsection{CL galaxy search}
\label{subsec:gal_samples}

To detect CLs in the BOSS LOWZ galaxy spectra, we used the detection algorithm outlined in detail in section 3.1 of \citeme\ and Clark et al. (in prep).
This algorithm primarily requires a galaxy to show at least three strong CLs to be classified as a CL galaxy (or two at $z>0.38$, where \shortfspecline{Fe}{xi} is no longer within the wavelength range of BOSS).
It also flags galaxies with at least one particularly strong CL with a signal-to-noise (SNR) $>10$ or at least two moderately strong \shortfspecline{Fe}{vii} lines with $\mathrm{SNR}>5$.
This search returned 419 CL galaxies.

\change{At certain redshifts, our detection algorithm misclassifies skylines as CLs redshifted to the same wavelengths.
In order to minimize this source of contamination, we construct redshift distributions of the LOWZ sample and the CL galaxies and visually inspect the CL galaxies in redshift bands that are over-represented compared to the overall galaxy sample.
We then remove all the galaxies in the redshift bands that are affected by skylines.}

\begin{figure}
	\centering
	\includegraphics[width=0.45\textwidth]{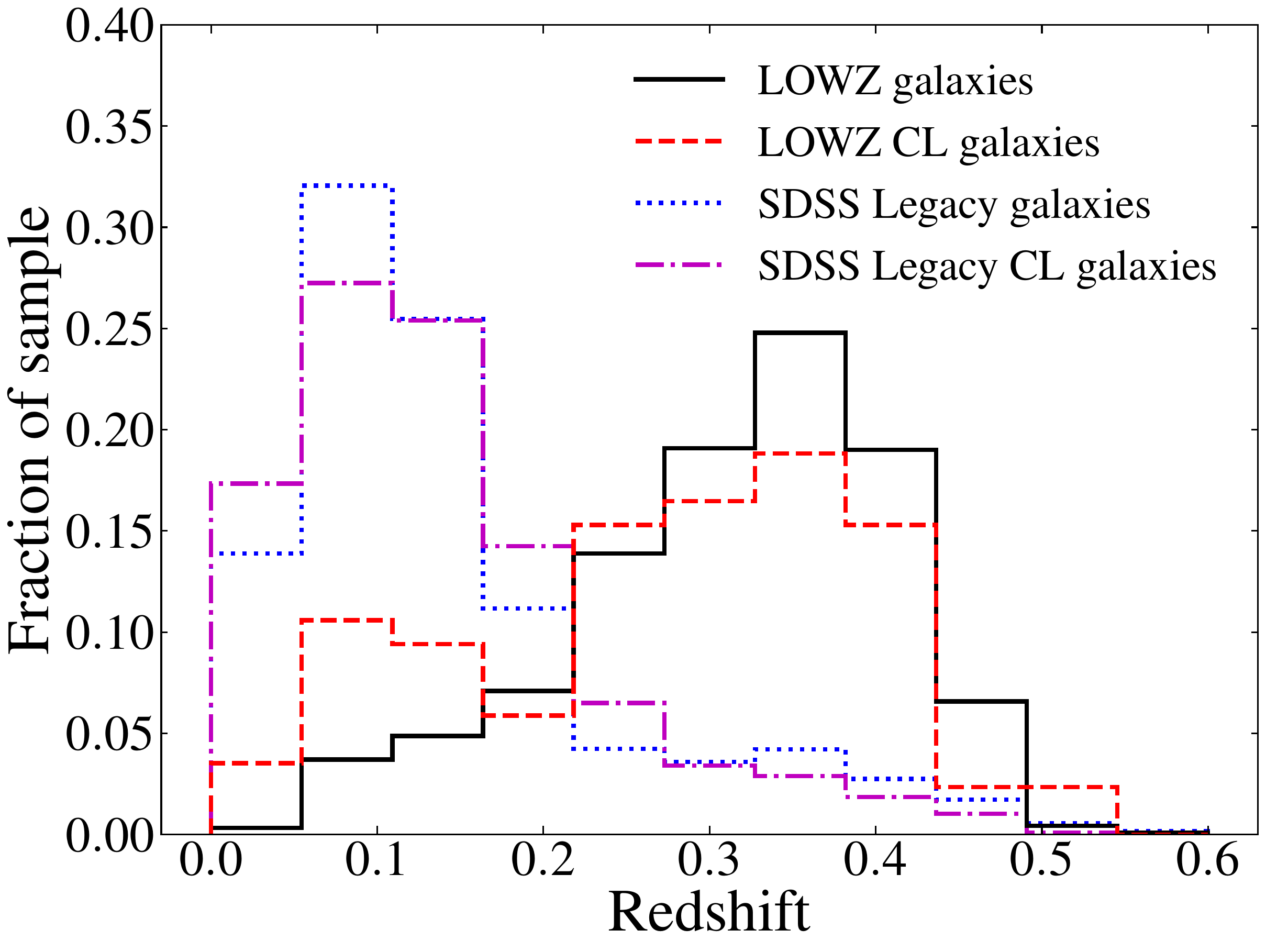}
	\caption{Redshift distributions of the LOWZ (solid black curve) and LOWZ CL (dashed red curve) galaxy samples, in comparison to the SDSS Legacy (dotted blue curve) and SDSS Legacy CL (dot dashed magenta curve) samples used by \citeme.
	The overall LOWZ and CL galaxy distributions are broadly consistent with each other. 
	\mycomment{with CL galaxies being under-represented between $0.1<z<0.18$ and $0.2<z<0.36$, and over-represented at $z\sim0.18$ and above $z\sim0.36$ at 99 per cent confidence.
	The over-representation occurs at redshifts where emission lines created by skylines are coincident with the observed wavelengths of \fspectralline{Fe}{vii}{6088}, \fspectralline{Fe}{x}{6376}, and \fspectralline{Fe}{xi}{7894}}}
	\label{fig:z_hist}
\end{figure}

We compared these CL galaxies to the overall LOWZ sample and the SDSS Legacy sample used by \citeme\ using their redshifts and total galactic stellar masses.
The LOWZ and SDSS Legacy samples both contain galaxies out to $z\sim0.6$, but distributed very differently (\figref{fig:z_hist}).
The median redshift of the Legacy sample is 0.11, compared to 0.38 for the LOWZ sample.
CL galaxies are detected across the full range of the LOWZ sample, but are not evenly distributed.
\mycomment{The over-representation of CL galaxies at redshifts $\sim0.18$, $\sim0.38$, and $\sim0.48$ is due to emission lines created by skylines being present at the observed wavelengths of \fspectralline{Fe}{vii}{6088}, \fspectralline{Fe}{x}{6376}, and \fspectralline{Fe}{xi}{7894}.}
\change{A Kolmogorov-Smirnov test between the LOWZ galaxy and CL galaxy samples produces a p-value $> 0.01$, which means we cannot reject the null hypothesis that the samples are drawn from the same population.}

\begin{figure}
	\centering
	\includegraphics[width=0.45\textwidth]{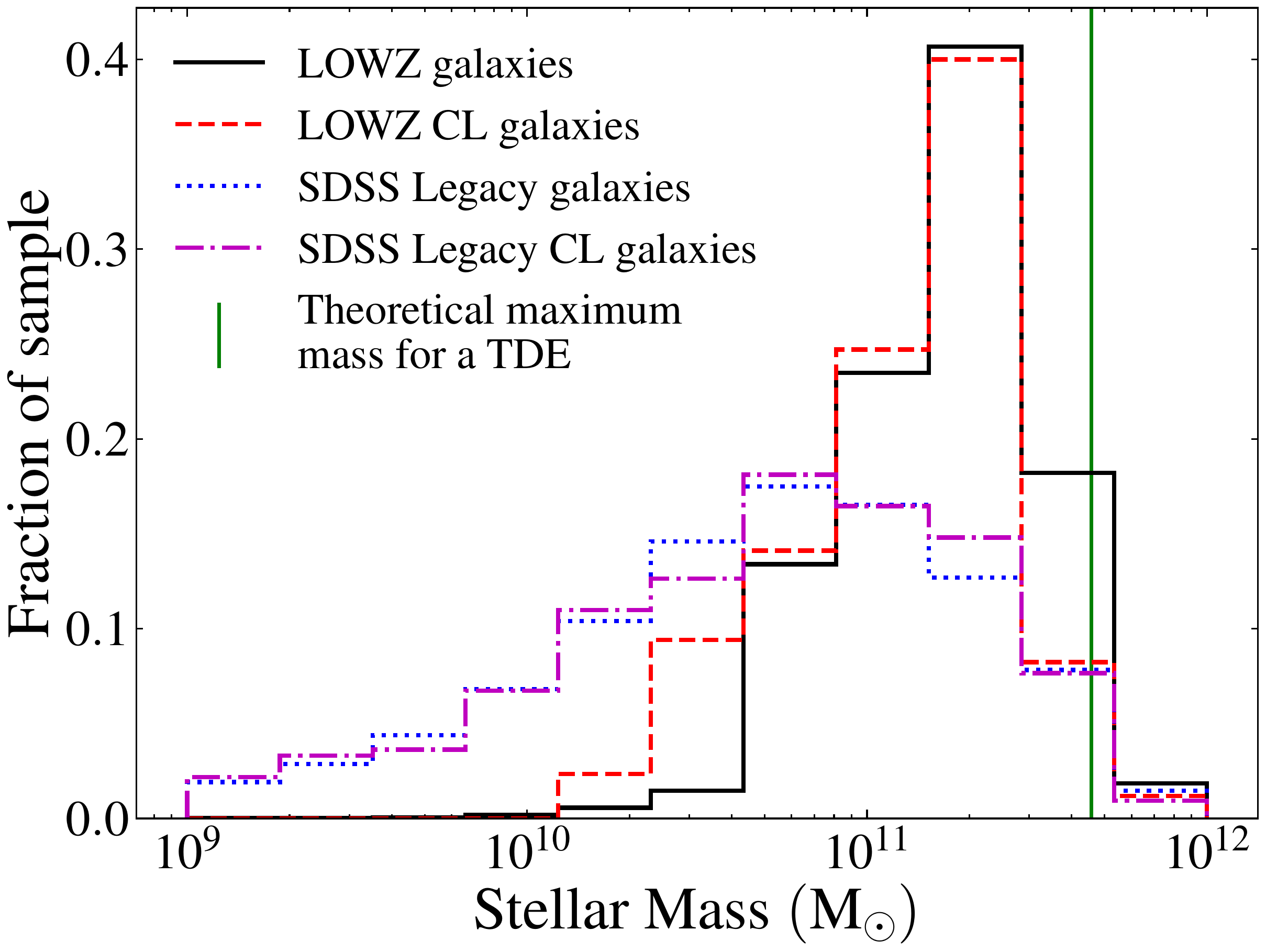}
	\caption{Galactic stellar mass distributions of the LOWZ (solid black curve) and LOWZ CL (dashed red curve) galaxy samples, in comparison to the SDSS Legacy (dotted blue curve) and SDSS Legacy CL (dot dashed magenta curve) galaxy samples used by \citeme.
	The vertical green line marks the theoretical galaxy stellar mass limit above which a Sun-like star would fall directly into the galaxy's SMBH instead of being disrupted as a TDE \citep{Rees1988}.
	The relation between the stellar mass of a galaxy and the mass of its SMBH used in this calculation is from \citet{Reines2015}.
	\change{The LOWZ distributions are broadly consistent but the CL galaxies are under-represented between $3\mathdash5\times10^{11}~\msun$ and over-represented below $5\times10^{10}~\msun$ at $>99$ per cent confidence.}}
	\label{fig:mass_hist}
\end{figure}

Compared to the SDSS Legacy sample used by \citeme, the LOWZ sample consists of galaxies with higher masses than most of the Legacy galaxies (\figref{fig:mass_hist}).
Therefore, by searching for ECLEs in LOWZ, we are probing a higher mass regime than in the SDSS Legacy Survey.
The mass distributions of the LOWZ sample and LOWZ CL galaxies have similar shapes with median masses of $1.8\times10^{11}$ and $1.5\times10^{11}~\msun$, respectively.
However, a two-sided Kolmogorov-Smirnov test allows us to reject the null hypothesis that the two distributions are drawn from the same population with a p-value $< 0.01$.
\change{This can be seen from the under-representation of CL galaxies with masses $3\mathdash5\times10^{11}~\msun$ and over-representation at the low mass ends of the distributions.}

\subsection{Detection efficiency}
\label{subsec:det_eff}

\begin{figure}
	\centering
	\includegraphics[width=0.45\textwidth]{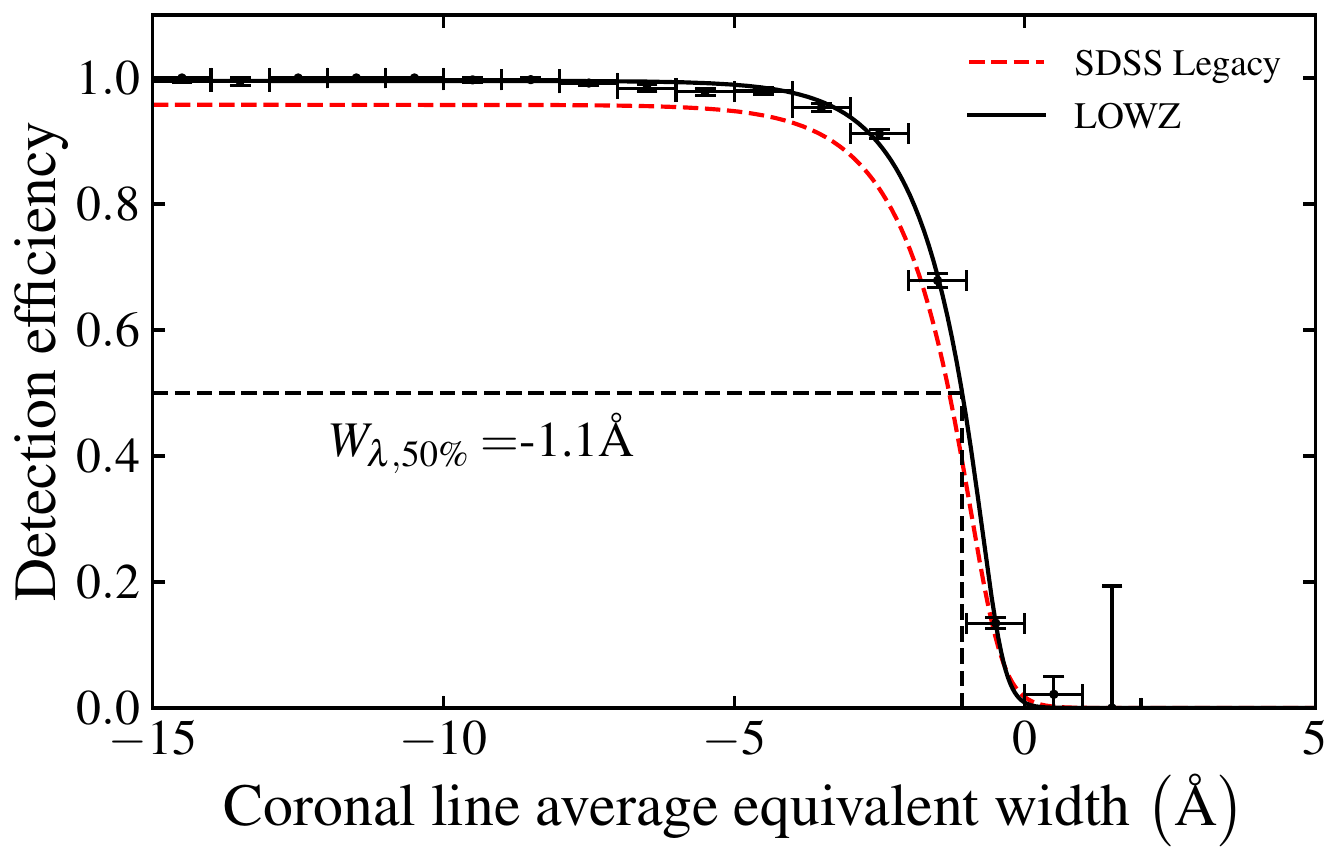}
	\caption{ECLE detection efficiency as a function of the average EW of the CLs.
    Points denote the fraction of fakes classified as ECLEs in 1 \AA\ bins.
    The black curve is a generalized sigmoid fit to the data.
    Error bars indicate 1$\upsigma$ binomial uncertainties.
	The red dashed curve is the efficiency of the detection algorithm when used on SDSS Legacy galaxies, as calculated by \citeme.}
	\label{fig:det_eff}
\end{figure}

Before deriving the rate at which vECLEs occur in BOSS LOWZ, we must first determine the efficiency of our detection algorithm.
We do this by running the algorithm on fake ECLE spectra created by planting CLs of varying strengths in LOWZ spectra.
10,000 spectra were randomly selected from the LOWZ sample described in \secref{subsec:boss_lowz}, into which the CLs from the \citet{Wang2012} ECLEs were planted.
For each base spectrum, the CLs' strengths were modified by a randomly selected scaling factor, which ranged from 0 to a maximum value that was set depending on the presence of \fspectralline{O}{iii}{5007} in the base spectrum.
If this feature was present, the maximum value was set such that the strongest CL would have the same strength as the \shortfspecline{O}{iii} line.
This was motivated by the fact that in the known ECLEs, the strengths of the CLs never exceeded the \shortfspecline{O}{iii} line.
If \shortfspecline{O}{iii} was not present, the maximum value \desichange{for the scaling factor} was set to one\desichange{, such that the scaled CLs could not be stronger than they were in the original ECLE spectrum}.
The results of running the detection algorithm on the fake LOWZ ECLEs is shown in \figref{fig:det_eff}.
The maximum detection efficiency is $\sim99$ per cent and we reach 50 per cent efficiency at an average equivalent width of $-1.1~\textrm{\AA}$.
This is very similar to the efficiency measured by \citeme\ when running the detection algorithm on SDSS Legacy DR17 galaxies.
For SDSS Legacy galaxies, the maximum efficiency was $\sim95$ per cent and 50 per cent efficiency was reached at an average equivalent width of $-1.3~\textrm{\AA}$.
This shows the algorithm is robust when analysing galaxies at a higher redshift.

\section{ECLE Sample}
\label{sec:ecle_sample}

In this section, we describe the cuts used to select the ECLEs from our sample of CL galaxies.
We then present the follow-up observations used to determine the transient nature of the ECLEs.

\subsection{ECLE criteria}
\label{subsec:ecle_criteria}

One of the key differences between the CLs in ECLEs and those typically produced by AGN are their strengths relative to the \fspectralline{O}{iii}{5007} emission line.
For AGN, the CL-to-\shortfspecline{O}{iii} ratio is a few per cent \citep{Nagao2000,Gelbord2009,Rose2015,Cerqueira-Campos2020}, whereas in ECLEs, it has been observed to be as high as 100 per cent \citep{Wang2012}.
We used the selection criteria from \citet{Wang2012}, which required at least one CL to be at least 20 per cent the strength of the \shortfspecline{O}{iii} line, and at least one CL to have an SNR $>5$.
These cuts reduced the sample to 293 galaxies.

We then visually inspected the remaining galaxies to remove false positives detected because of random noise or coincidence of skylines with the locations of CLs.
This process returned two ECLEs, the details of which are summarized in \tabref{tab:sample}.

\begin{table*}
	\centering
	\caption{Summary of the ECLEs detected in the BOSS LOWZ sample.}
	\begin{tabular}{|c|c|c|c|c|}
		\hline
		SDSS Name & Short Name & RA (J2000) & Dec (J2000) & Redshift \\
		\hline
		SDSS J011306.68+093712.2 & SDSS J0113 & 01:13:06.6834 & +09:37:12.2716 & 0.1537 \\
		SDSS J221831.51+233432.3 & SDSS J2218 & 22:18:31.5153 & +23:34:32.3013 & 0.0787 \\
		\hline
	\end{tabular}
	\label{tab:sample}
\end{table*}

As we do not perform full spectral fitting when measuring the emission line strengths to identify CL galaxies and ECLEs, we may not be sensitive to line enhancements due to \shortfspecline{Fe}{ii} complexes or the blending of adjacent features.
This may affect whether CLs appear to be stronger than the threshold that we use to select ECLEs.
To investigate this, we check for CL galaxies with a CL-to-\shortfspecline{O}{iii} ratio between 10 and 20 per cent.
We find no CL galaxies that meet this criterion, which suggests that this is a suitable to cut to select ECLEs.

\subsection{Transient nature}
\label{subsec:transient}

To search for transient activity in the new ECLEs, we performed a cross-match of the new ECLEs with the Transient Name Server to search for previous transient activity.\footnote{https://www.wis-tns.org/}
Neither of the ECLEs showed any previous activity, supporting our assumption that if the CLs were created by a transient progenitor, it was not a recurring process.

In order to determine the transient nature of the ECLEs, we used follow-up optical spectra (\figref{fig:spec_comp}) and MIR photometric observations (\figref{fig:mir_summary}).
We obtained optical spectra of both galaxies; SDSS J0113 using GMOS-N and SDSS J2218 using DESI.
Given that there was a gap of $\sim10$ yr between the BOSS spectra and our follow-up observations, if the ECLEs were variable, we would expect the CLs to have faded significantly between the two sets of observations.
vECLEs also exhibit a distinctive evolution when viewed in the MIR, particularly when looking at the $W1-W2$ colour of the objects.
\citet{Clark2024} and \citet{Hinkle2024} found that the vECLEs all showed long term declines in the MIR, whereas the non-vECLEs remained constant in their MIR emission.
Additionally, \citet{Clark2024} observed that the $W1-W2$ colour of the vECLEs evolved from above the $W1-W2>0.8$ mag AGN cut \citep{Stern2012} to well below this cut.
\citet{Clark2024} concluded that this colour index evolution is a robust method to select vECLEs.

\subsubsection{SDSS J0113+0937}
\label{subsubsec:J0113}

SDSS J0113 was detected due to having several moderately-strong \shortfspecline{Fe}{vii} lines, but has no other CLs.
It appeared to show weak \shortfspecline{Fe}{xi} emission, but visual inspection revealed this to be skylines coincident with the \shortfspecline{Fe}{xi} observed wavelength.
The \shortfspecline{Fe}{vii} lines have disappeared completely in the GMOS spectrum, strongly suggesting that this object is a vECLE.
However, in contrast to the vECLE sample, the \fspectralline{O}{iii}{5007} line has also decreased, whereas in the known vECLEs, this line remained roughly constant or increased in strength over a similar time period between observations.
The LOWZ spectrum also had prominent \Ha\ emission and weak \spectralline{He}{ii}{4686}.
These emission lines have all changed in the GMOS spectrum, with the \Ha\ and \shortspecline{He}{ii} emission both weakening significantly.
The \Ha\ complex is affected by telluric absorption in the GMOS spectrum, so we are unable to confidently determine the extent to which it has weakened.

\mnraschange{We note the presence of strong \fspectralline{Ne}{v}{3426} emission in the BOSS spectrum of SDSS J0113, which has disappeared by the time of the GMOS spectrum.
This is one of a doublet of emission lines, the other being \fspectralline{Ne}{v}{3347} which is not detected in SDSS J0113.
\fspectralline{Ne}{v}{3426} (hereafter \shortfspecline{Ne}{v}) has been previously noted as a high-ionization CL primarily present in AGN spectra \citep{Abel2008,Gilli2010,Mignoli2013,Cleri2023a,Cleri2023b}, although stellar light from Wolf-Rayet stars and shocks from energetic SNe have been proposed as alternate production mechanisms \citep{Izotov2012,Zeimann2014,Olivier2022}.
\shortfspecline{Ne}{v} has a very similar ionizing potential to \fspectralline{Fe}{vii}{6088} ($\sim100$ eV), so it is not surprising that it is present in the spectrum of SDSS J0113.
However, \shortfspecline{Ne}{v} was not observed in any of the ECLEs detected in the SDSS Legacy survey, as they were all at low redshifts where the emission line fell outside the wavelength range of the Legacy spectrograph.
As SDSS J0113 is at a higher redshift than the Legacy ECLEs and the wavelength range of the BOSS spectrograph is broader than for the Legacy survey, we are able to observe this emission line in an ECLE for the first time.
The \shortfspecline{Ne}{v} emission line is five times stronger than the \fspectralline{Fe}{vii}{6088} emission line and $2.5$ times stronger than the \fspectralline{O}{iii}{5007} emission line, making it one of the strongest emission lines in the spectrum.
As \shortfspecline{Ne}{v} was not observed in the SDSS Legacy ECLE sample, we do not use this line when performing the vECLE rate analysis.
Given the iron emission lines were the focus of that sample, our assumptions and modelling are based on the properties of the iron lines in the Legacy ECLEs.
We do not know if the \shortfspecline{Ne}{v} line evolves in the same way as the iron lines, so we are unable to apply our modelling to this line.
In addition, \shortfspecline{Ne}{v} lay outside the wavelength range of the observations of AT~2017gge, the TDE with CLs that we use in our rate analysis.
However, \shortfspecline{Ne}{v} has been detected in the spectra of some of the TDEs that have developed CLs, namely AT~2018dyk \citep{Frederick2019,Clark2025}, AT~2019aalc \citep{Veres2024}, and TDE~2019qiz \citep{Short2023}.
Therefore, though we are unable to use this emission line in this work, it will be useful in future ECLE searches at higher redshifts, where \shortfspecline{Ne}{v} will more often fall within the observed wavelength range.}

The MIR behaviour of SDSS J0113 closely aligns with that of the vECLE sample.
Both bands show a decline over a timescale of years, followed by a slower decline (or, perhaps, a plateau).
Both bands also show an increase in magnitude at the start of the observation period, indicating that the peak of the MIR emission occurred at a MJD of $\sim55300$.

The colour index also shows similar evolution to that of vECLEs, evolving from an AGN-like colour to a non-AGN-like colour.
The first two points of the colour index evolution show the colour evolving in the direction of the AGN-like colour space, indicating that the pre-outburst colour was also non-AGN-like.
Interestingly, the peak in colour appears to lag the individual bands' peaks by $\sim700$ d.
However, due to the gap in \textit{WISE} observations, we are unable to constrain this lag.

To further investigate the MIR behaviour, we fit each band's decline using a power law of the same form as expected for TDE light curves.
In \figref{fig:index_comp}, we show the power law index of the $W1$ and $W2$ declines for SDSS J0113 and compare them to the vECLE sample and X-ray selected TDEs.
The fits to each band give a very similar power law index, with values of 0.43 and 0.40 for $W1$ and $W2$, respectively.
The weighted mean of the indices sits within the range of TDE indices from \citet{Auchettl2017} and is very close to the expected index from the disk emission model.
Both the spectral and MIR evolution of SDSS J0113 are consistent with the known vECLEs.
Therefore, we classify it as a vECLE.

\subsubsection{SDSS J2218+2334}
\label{subsubsec:J2218}

SDSS J2218 was selected as an ECLE due to the strength of the \fspectralline{Fe}{vii}{6088} line.
It also exhibited prominent \Ha\ and weak \fspectralline{O}{iii}{5007} emission.
However, it did not show any other CLs\mnraschange{, including \fspectralline{Ne}{v}{3426}}.
The DESI spectrum is very similar to the LOWZ spectrum.
The shape of the continuum has not changed, and the strength of the emission lines has remained unchanged.

The MIR light curve also indicates this object is not transient, as the emission in both bands has remained roughly constant, as has the colour index.
Therefore, we are confident in classifying SDSS J2218 as non-variable and therefore not a vECLE.

We note, however, that both the spectrum and MIR colour index of this object are very different in comparison to the previously described non-vECLEs \citep{Wang2012,Yang2013,Clark2024,Hinkle2024}.
Previous non-vECLE spectra were all consistent with AGN or star-forming galaxies, with a predominantly blue continuum and broad Balmer features.
SDSS J2218, instead, has a red continuum and narrow \Ha\ emission.
This difference is also seen in the MIR colour index, where SDSS J2218 lies below the AGN dividing line, whereas all the other non-vECLEs sit above this line \citep[][Fig. 5]{Clark2024}.
The spectrum's continuum and emission lines are consistent with a galaxy dominated by post-AGB star emission, which produces the red spectrum with narrow \Ha\ and little \shortfspecline{O}{iii} emission \citep{Habing1997}.

\begin{figure*}
	\centering
	\includegraphics[width=\textwidth]{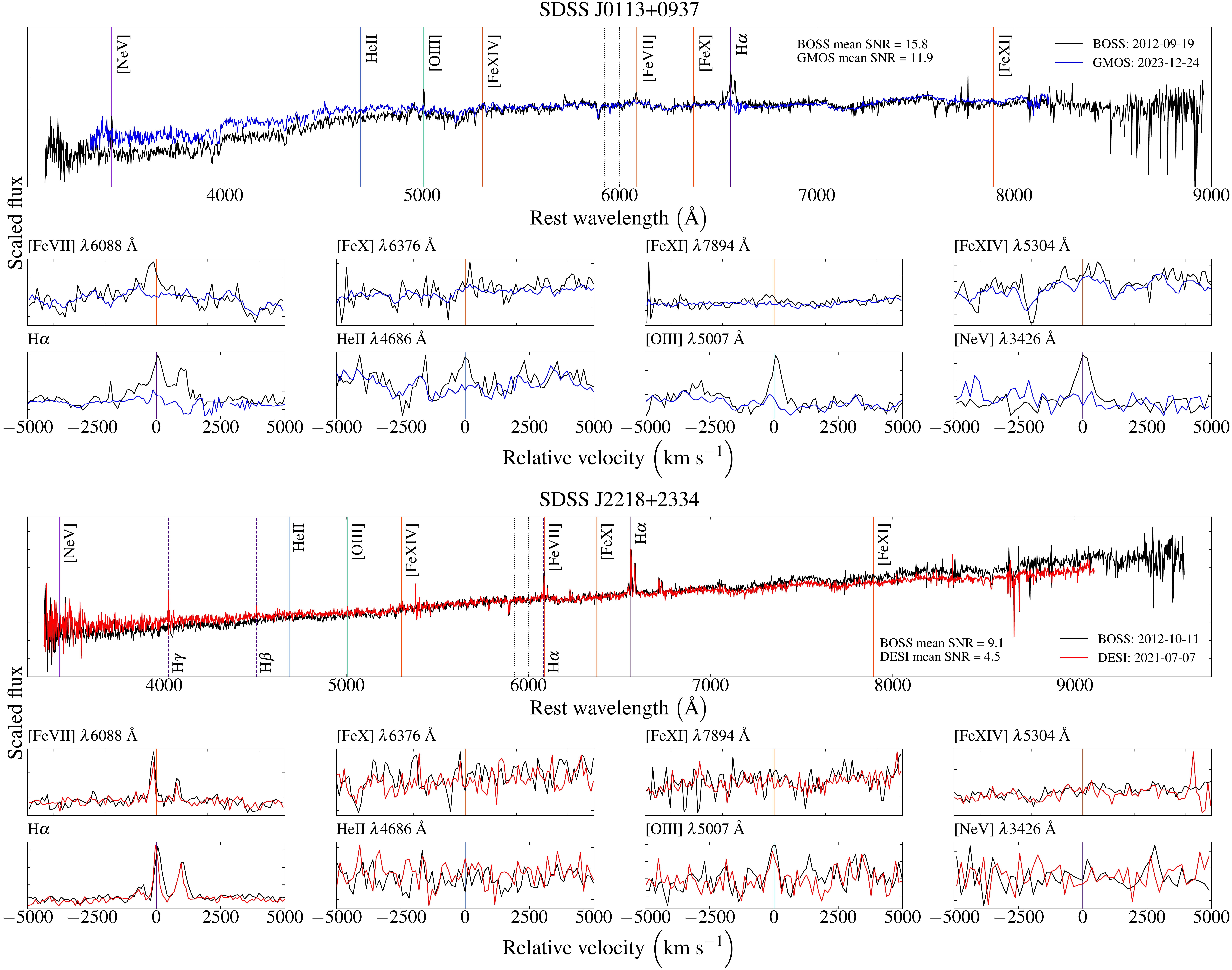}
	\caption{Comparisons of the ECLE spectra from BOSS LOWZ (black), GMOS (blue) and DESI (red).
	Emission lines of interest are marked by vertical solid lines with labels at the top of the plots.
	The dotted lines in the upper plots indicate the region used to rescale the spectra with respect to each other to allow for easy comparison.
	For the line-specific plots, this rescaling was done on continuum sections of the spectra near the emission line.
	For SDSS J2218, the vertical dashed lines indicate the $z\simeq0$ Balmer emission that is imposed on top of the galaxy spectrum.
	The labels for these lines are at the bottom of the plot.}
	\label{fig:spec_comp}
\end{figure*}

\begin{figure*}
	\centering
	\includegraphics[width=\textwidth]{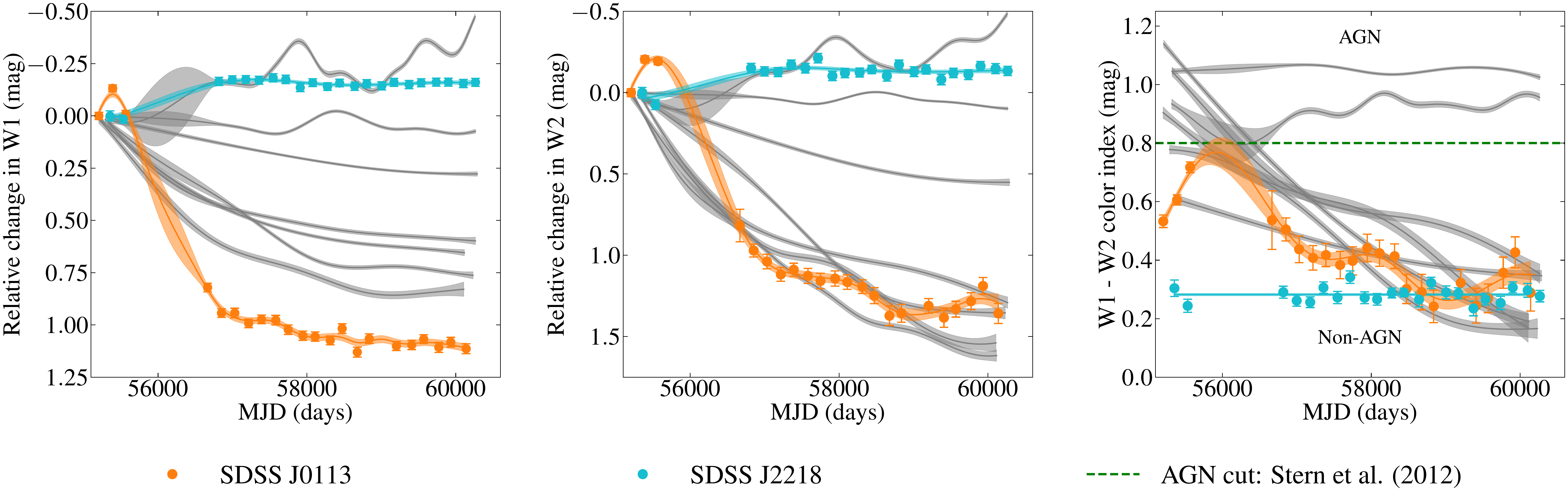}
	\caption{Comparison of the MIR evolutions of SDSS J0113 (orange) and SDSS J2218 (blue) with the \citet{Wang2012} ECLE sample (grey).
	The shaded region for each object is the $1\upsigma$ uncertainty on the best fit, determined using a Gaussian process.
	The left and centre panels show the evolution in the W1 and W2 bands.
	The right panel shows the colour evolution.
	The dashed horizontal line is the AGN/non-AGN dividing line from \citet{Stern2012}.
	SDSS J2218 shows little evolution in the $W1$ and $W2$ bands and no colour evolution, so is inconsistent with the vECLEs.
	SDSS J0113 shows a clear decline between the \textit{WISE} and \textit{NEOWISE} observations periods in a manner consistent with the vECLE sample.
	This decline is also seen in the colour plot, again consistent with the vECLEs.
	}
	\label{fig:mir_summary}
\end{figure*}

\begin{figure*}
	\centering
	\includegraphics[width=\textwidth]{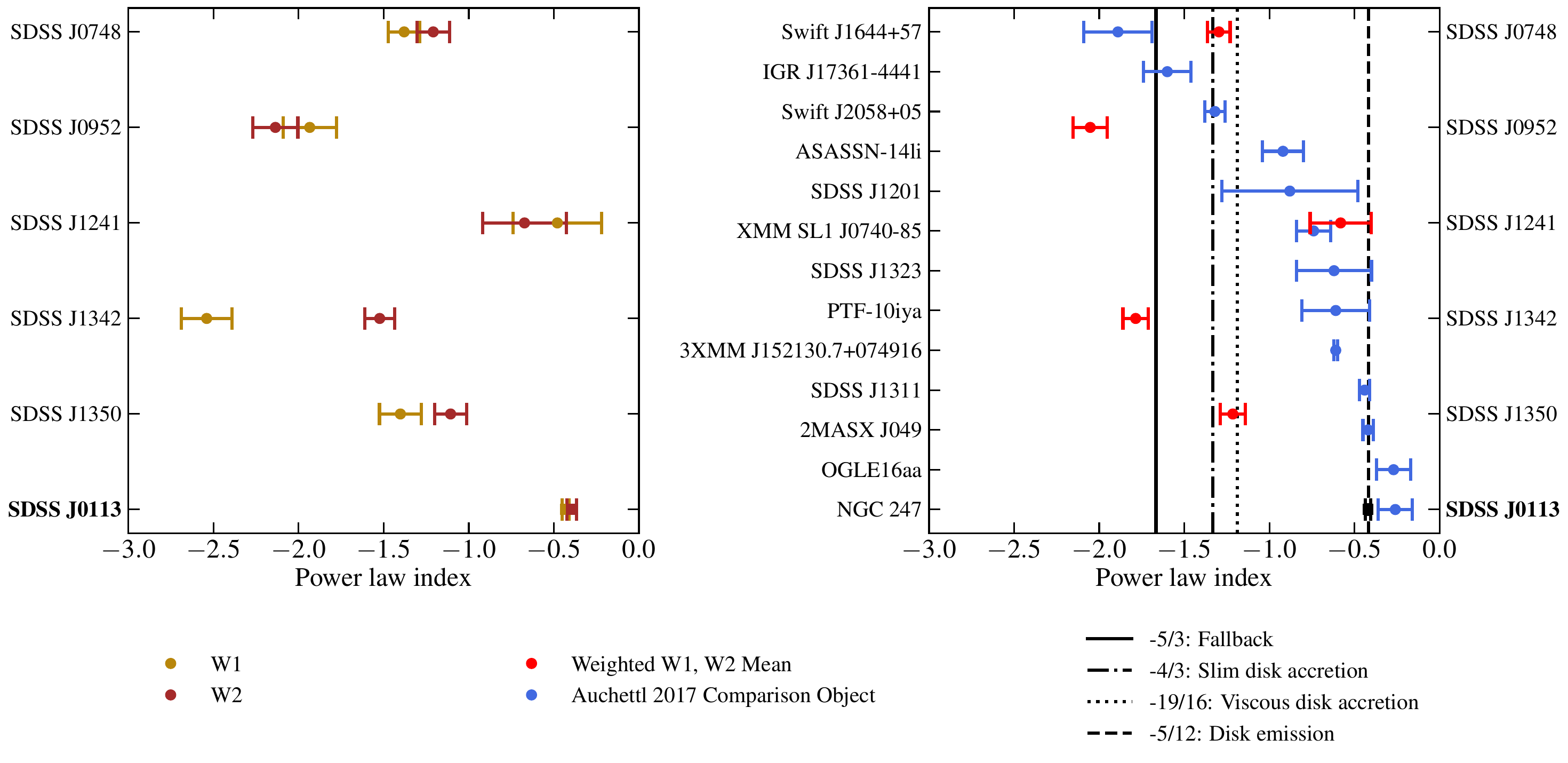}
	\caption{Comparison of the power law indices fit to the declining sections of the MIR light curves of the literature vECLE sample and SDSS J0113, and power law indices of X-ray selected TDE light curves from \citet{Auchettl2017}.
	\textit{Left}: Power law indices for the W1 and W2 band fits for the five known vECLEs and the possible vECLE from this work, SDSS J0113.
	\textit{Right}: Weighted mean power law indices for the vECLEs and power law indices for the X-ray TDEs from \citet{Auchettl2017}.
	SDSS J0113 is represented by square markers in both plots.
	Both bands give similar indices for SDSS J0113.
	The weighted mean lies within the range of indices from the \citet{Auchettl2017} TDEs and is close to the $-\frac{5}{12}$ index from the disk emission model.}
	\label{fig:index_comp}
\end{figure*}

Upon investigation of the \fspectralline{Fe}{vii}{6088} emission line, it was found that this line and other emission lines in the spectrum were consistent with Balmer emission at $z\simeq0$.
The \Ha\ emission line is coincident with the observed wavelength of \fspectralline{Fe}{vii}{6088}.
These lines are marked on \figref{fig:spec_comp}.
The galaxy lies in a higher density region of the \citet{Finkbeiner2003} \Ha\ sky-map, and nearby galaxies exhibit the same $z\simeq0$ Balmer emission.
As there is no intervening object visible, we conclude that this Balmer emission was created by a diffuse cloud of hydrogen in the Milky Way.
Therefore, this object is a false positive and not included in the rate calculations.

\section{ECLE Rate Analysis}
\label{sec:rate_analysis}

In this section, we calculate the galaxy-normalized, mass-normalized, and volumetric vECLE rates in the BOSS LOWZ sample using the one vECLE described above.
The process for calculating the rates is as follows.
First, we calculate the total visibility time of our survey (\secref{subsec:vis_time}).
We then calculate a galaxy-normalized vECLE rate (\secref{subsec:gal_rate}) and a mass-normalized rate, where the visibility time of each galaxy in our sample is weighted by its stellar mass (\secref{subsec:mass_rate}).
Next, we convert the mass-normalized rate to a volumetric rate using a galactic stellar mass function (\secref{subsec:vol_rate}).
Finally, we compare the LOWZ vECLE rates to the SDSS Legacy vECLE rates and TDE rates from the literature (\secref{subsec:rate_comp}).

\subsection{Visibility time}
\label{subsec:vis_time}

The visibility time for each galaxy is the amount of time a vECLE would be detectable at that galaxy's redshift.
Here, we follow the same process as \citeme.
First, we model the strengths of the CLs over time.
This entails sampling a random peak luminosity using the TDE X-ray luminosity function (LF) from \citet{Sazonov2021}, converting it to a peak CL strength using observations of the CL-TDE AT~2017gge \citep{Onori2022}, and then evolving this strength over the period of 10 yr according to a power law, which is the predicted shape of a TDE light curve.
This length of time was chosen as the CLs in the \citet{Wang2012} vECLE sample had all significantly weakened or disappeared this long after their SDSS discovery spectra.

The \citet{Sazonov2021} TDE LF is parametrized as
\begin{equation}
	\frac{\mathrm{d}\dot{N}}{\mathrm{d}\log_{10}L}=\dot{N}_0(L/L_0)^a,
	\label{eq:vv_lf}
\end{equation}
where $\dot{N}$ is the volumetric rate and $L$ is the peak luminosity of the TDE.
The constants are $L_0=10^{43}\ \mathrm{erg\ s^{-1}}$, $\dot{N}_0=(1.4\pm0.8)\times10^{-7}\ \mathrm{Mpc^{-3}~yr^{-1}}$ and $a=-0.6\pm0.2$.
As the LF gives the rate of TDEs of a particular peak luminosity, we normalized it over the range of luminosities the LF was constructed over, $10^{42.7}-10^{44.9}\ \mathrm{erg\ s^{-1}}$, to create a probability distribution.

For each galaxy in the LOWZ survey, we sample a peak luminosity, $L_\mathrm{max}$, and convert it into a peak CL strength, $S_\mathrm{max}$, by requiring that the ratio between the peak CL strength and AT~2017gge's peak CL strength, $S_\mathrm{gge}$, at 218 d post-discovery is the same as the ratio of the selected peak luminosity and AT~2017gge's peak X-ray luminosity, $L_\mathrm{gge}$, i.e.,
\begin{equation}
	\frac{S_\mathrm{max}}{S_\mathrm{gge}}=\frac{L_\mathrm{max}}{L_\mathrm{gge}}.
	\label{eq:peaks}
\end{equation}
Here we assume that a TDE with a larger peak X-ray luminosity will produce CLs with a higher peak strength.
The process used to measure the peak CL strengths of AT~2017gge is described by \citeme.

The peak CL strength is evolved as the power law
\begin{equation}
	S=\alpha t^{\beta},
	\label{eq:power}
\end{equation}
where $S$ is the average strength of the CLs, $t$ is the time since the peak CL strength, and $\alpha$ and $\beta$ are constants.
Theoretical models have predicted power law indices between $-5/12$ and $-5/3$ (\figref{fig:index_comp}).
Therefore, for each galaxy we sample the index from this range assuming a flat probability density function.
Observational studies have found a larger range of indices \citep{Auchettl2017}, which we use when determining the uncertainty on the rate.

Next, the CL strength curve is redshifted according to the galaxy's redshift, which, due to cosmological time dilation, increases the time over which the CLs are visible.
The visibility time, $t_\mathrm{v}$, for each galaxy is then
\begin{equation}
	t_\mathrm{v}=\int\epsilon[S(t)]\mathrm{d}t,
	\label{eq:vis_time}
\end{equation}
where $\epsilon(S)$ is the detection efficiency as a function of CL strength measured in \secref{subsec:det_eff} and the integral runs over the full time over which the strength evolution is modelled.

\subsection{Galaxy-normalized rate}
\label{subsec:gal_rate}
The galaxy-normalized vECLE rate is the number of vECLEs discovered in the BOSS LOWZ sample divided by the sum of the visibility times of all the galaxies searched over,
\begin{equation}
	R_\mathrm{G}=\frac{N_{\mathrm{ECLE}}}{\sum_{i=1}^{N_\mathrm{g}}t_{\mathrm{v},i}},
	\label{eq:gal_rate}
\end{equation}
where $N_{\mathrm{ECLE}}$ is the number of vECLEs detected, $N_\mathrm{g}$ is the number of galaxies searched over, and $t_{\mathrm{v},i}$ is the visibility time of the \textit{i}-th galaxy.

Due to the small number of vECLEs detected, the dominant source of statistical error in our rate calculations is the Poisson uncertainty on our single detection.
This gives an error on the number of ECLEs detected of $N_\mathrm{ECLE}=1.0~^{+2.3}_{-0.8}$.
The statistical uncertainty on the total visibility time stems from the uncertainties on the average peak strength of the CLs of AT~2017gge, the peak X-ray luminosity of AT~2017gge, and the range of power law indices used to evolve the CL strengths. 
To determine how these individual uncertainties propagate to the uncertainty on the total visibility time, we perform a Monte Carlo simulation, which entails calculating the total visibility time 500 times, randomly sampling the parameters from their probability density distributions each time.
The AT~2017gge peak luminosity and line strength are drawn from normal distributions with standard deviations set to their $1\upsigma$ errors.
For the power law indices, we vary the range from which the indices are sampled using the observational range from \citet{Auchettl2017}, who found power law indices ranging from $-0.26\pm0.10$ to $-1.89\pm0.20$.
Having calculated the uncertainties on the number of ECLEs detected and the total visibility time, we use standard error propagation to determine the uncertainty on the galaxy-normalized rate.
This yields a galaxy-normalized rate of \lowzgalrate.

As BOSS observed higher-mass galaxies on average than SDSS Legacy (see \figref{fig:mass_hist}), we are able to extend the vECLE rate vs. mass relation found by \citeme\ to higher masses.
\desichange{\citeme\ detected five vECLEs in SDSS Legacy, so the galaxies in that sample were put into bins based on the masses of the vECLE host galaxies.
The vECLE rate was then calculated for each bin individually, where the lowest and highest mass bins contained no vECLEs and were therefore used as upper limits on the rate for zero detections.
As we only detected one vECLE in LOWZ, the galaxy-normalized vECLE rate from the LOWZ sample is added to the relation at the median mass of the LOWZ sample rather than binning the galaxies.}
The stellar masses and rates of the SDSS Legacy bins and the LOWZ rate are shown in \tabref{tab:rate_mass_relations}.

\begin{table}
	\centering
	\caption{vECLE galaxy and mass-normalized rates used in fitting the rate-mass relations.}
	\begin{tabular}{ccc}
		\hline
		Stellar mass & Galaxy-normalized rate & Mass-normalized rate \\
		$\left(10^{10}\msun\right)$ & $\left(10^{-6}~\mathrm{galaxy}^{-1}~\mathrm{yr}^{-1}\right)$ & $\left(10^{-17}~\mathrm{M_\odot}^{-1}~\mathrm{yr}^{-1}\right)$ \\
		\hline
		\multicolumn{3}{c}{\textbf{SDSS Legacy}} \\
		$0.13~^{+0.12}_{-0.09}$ & $<40$ & $<2200$ \\ [1mm]
		$0.7~^{+0.3}_{-0.3}$ & $13~^{+9}_{-5}$ & $160~^{+110}_{-60}$ \\ [1mm]
		$2.5~^{+1.2}_{-0.9}$ & $5.6~^{+4.0}_{-2.2}$ & $19~^{+13}_{-7}$ \\ [1mm]
		$8.0~^{+4.5}_{-2.8}$ & $2~^{+5}_{-2}$ & $2~^{+5}_{-2}$ \\ [1mm]
		$26~^{+15}_{-8}$ & $<13$ & $<3.7$ \\
		\multicolumn{3}{c}{\textbf{BOSS LOWZ}} \\
		$18~^{+10}_{-13}$ & $1.6~^{+3.8}_{-1.4}$ & $0.7~^{+1.6}_{-0.6}$ \\
		\hline
		\multicolumn{3}{l}{The rates marked with $<$ represent $2\upsigma$ upper limits.}
	\end{tabular}
	\label{tab:rate_mass_relations}
\end{table}

\begin{figure}
	\centering
	\includegraphics[width=0.45\textwidth]{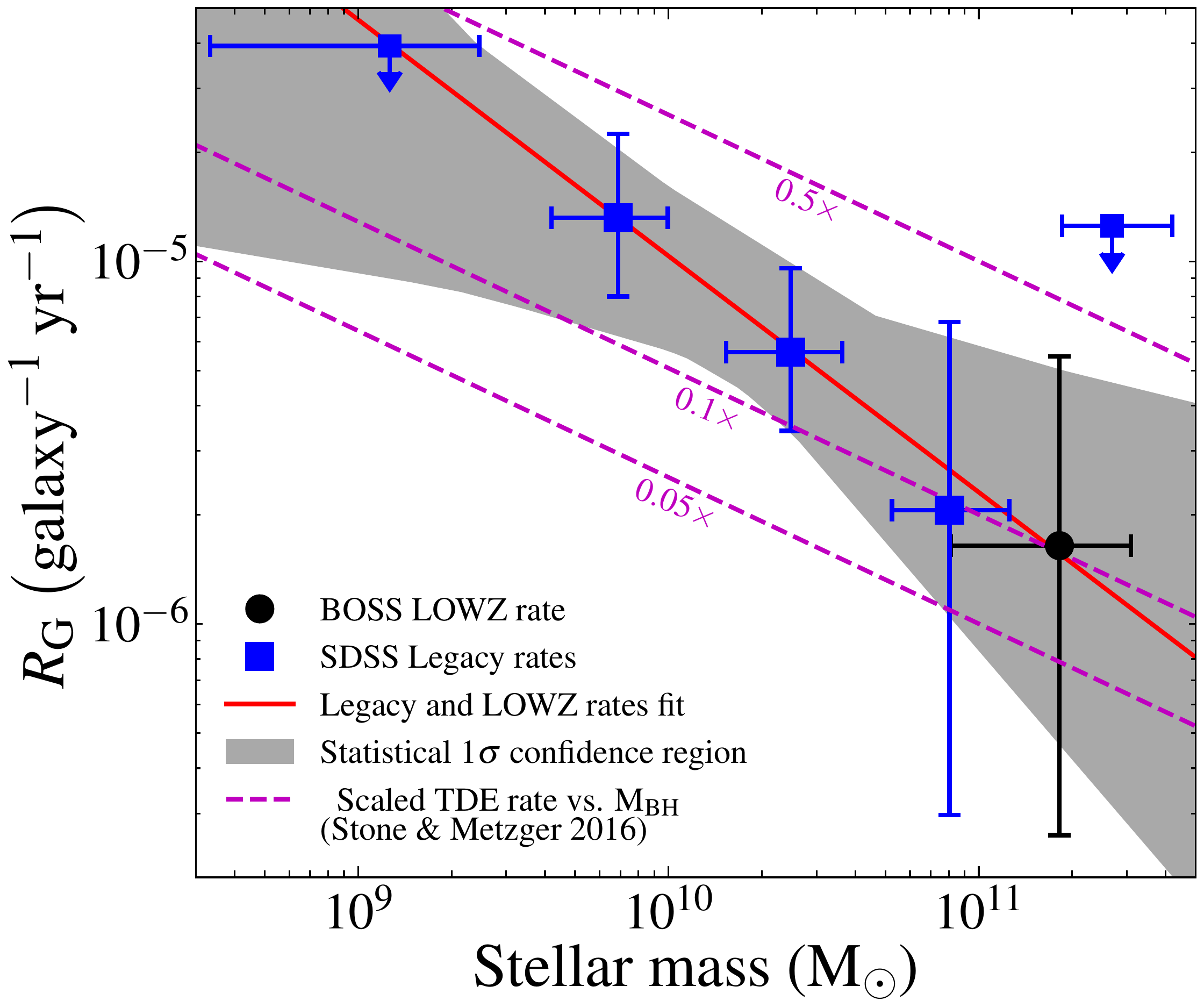}
	\caption{Galaxy-normalized vECLE rates as a function of galaxy stellar mass for SDSS Legacy (blue squares) and BOSS LOWZ (black circle).
	Vertical error bars show the statistical errors on the rates derived using the Monte Carlo simulations detailed above; and the horizontal error bars denote the range within each mass bin that 68 per cent of the galaxies fall.
	The points marked with downward arrows are $2\upsigma$ upper bounds on the rates calculated using the upper Poisson error on zero detections.
	The solid red line shows the power law fit to the Legacy and LOWZ rates and the shaded area is the $1\upsigma$ confidence region.
	The dashed purple line shows the TDE rate vs. black hole mass relation calculated by \citet{Stone2016}, scaled by 0.05, 0.1, and 0.5.}
	\label{fig:gal_rate_mass_relation}
\end{figure}

The rate vs. mass relation is then refit using the SDSS Legacy and LOWZ points with a power law of the form
\begin{equation}
	\log_{10}(R_\mathrm{M})=a\log_{10}(M)+b.
	\label{eq:rate_mass_relation}
\end{equation}
\desichange{To estimate the fit of this relation, we calculate the $\upchi^2$ for 1,000,000 fits using pairs of parameters with values between $-3<a<0$ and $-10<b<10$ and take the pair that produces the lowest $\upchi^2$ as the best fit.
We use the rates of the three SDSS Legacy bins that contain vECLEs and the BOSS LOWZ rate to perform this fit.
As the error bars on the rates are asymmetric, when calculating the $\upchi^2$ of each fit, we use the positive error when the residual is positive, and the negative error when the residual is negative.
The upper limits on zero detections from SDSS Legacy are implemented by requiring that the fits calculated from the pairs of parameters must fall below the upper limits in their mass ranges.}
This fit (shown in \figref{fig:gal_rate_mass_relation}) estimates values of $a=-0.7\pm0.3$ and $b=1.5^{+2.9}_{-3.5}$, with a reduced $\upchi^2=0.02$.
This fit is consistent with the one measured by \citeme.
\desichange{Though $\sim30$ per cent of galaxies in the LOWZ sample were also observed by the SDSS Legacy survey, there is no overlap between the CL galaxy samples drawn from each survey.
Therefore, the same vECLE was not detected in both samples and so the rates are independent.}

We compare the rate-mass relation in \figref{fig:gal_rate_mass_relation} to the theoretical TDE rate vs. black hole mass relation calculated by \citet{Stone2016} by converting black hole mass to galactic stellar mass using the relation from \citet{Reines2015}.
The dashed purple lines in \figref{fig:gal_rate_mass_relation} show this relation, scaled by factors of 0.05, 0.1, and 0.5 to allow better comparison to the vECLE rates.
The two relations have similar shapes (within their measured uncertainties) and the $1\upsigma$ confidence region on our measured relation lies between the 10 and 50 per cent scalings of the TDE rate relation.
This is consistent with the findings of \citeme\ and supports our suggestions that vECLEs are caused by a subset of TDEs.

We caution that the $\upchi^2$ distribution used to determine the power-law fit parameters and confidence region is a biased estimator due to the inclusion of Poisson uncertainties stemming from the small number of vECLEs per bin. Hence, the errors on the parameters and the confidence region are likely underestimated. However, we note that the shape of the power-law fit and the theoretical TDE rate vs. mass relation are still qualitatively consistent.

\subsection{Mass-Normalized Rate}
\label{subsec:mass_rate}
We calculate the mass-normalized ECLE rate by weighting the visibility time of each galaxy by its stellar mass, $M_\star$, in the rate calculation, i.e.,
\begin{equation}
	R_\mathrm{M}=\frac{N_{\mathrm{ECLE}}}{\sum_{i=1}^{N_\mathrm{g}}t_{\mathrm{v},i}M_{\star,i}}.
	\label{eq:mass_rate}
\end{equation}
We use the masses derived by the Portsmouth group pipeline, as described in \secref{subsec:boss_lowz}.
We repeat the Monte Carlo simulation to determine the statistical uncertainty on the rate, with the addition of varying the masses according to their uncertainties.
To that end, the masses are drawn from normal distributions with the standard deviation set to the $1\upsigma$ errors on the masses.
The resultant mass-normalized rate is \lowzmassrate.

\begin{figure}
	\centering
	\includegraphics[width=0.45\textwidth]{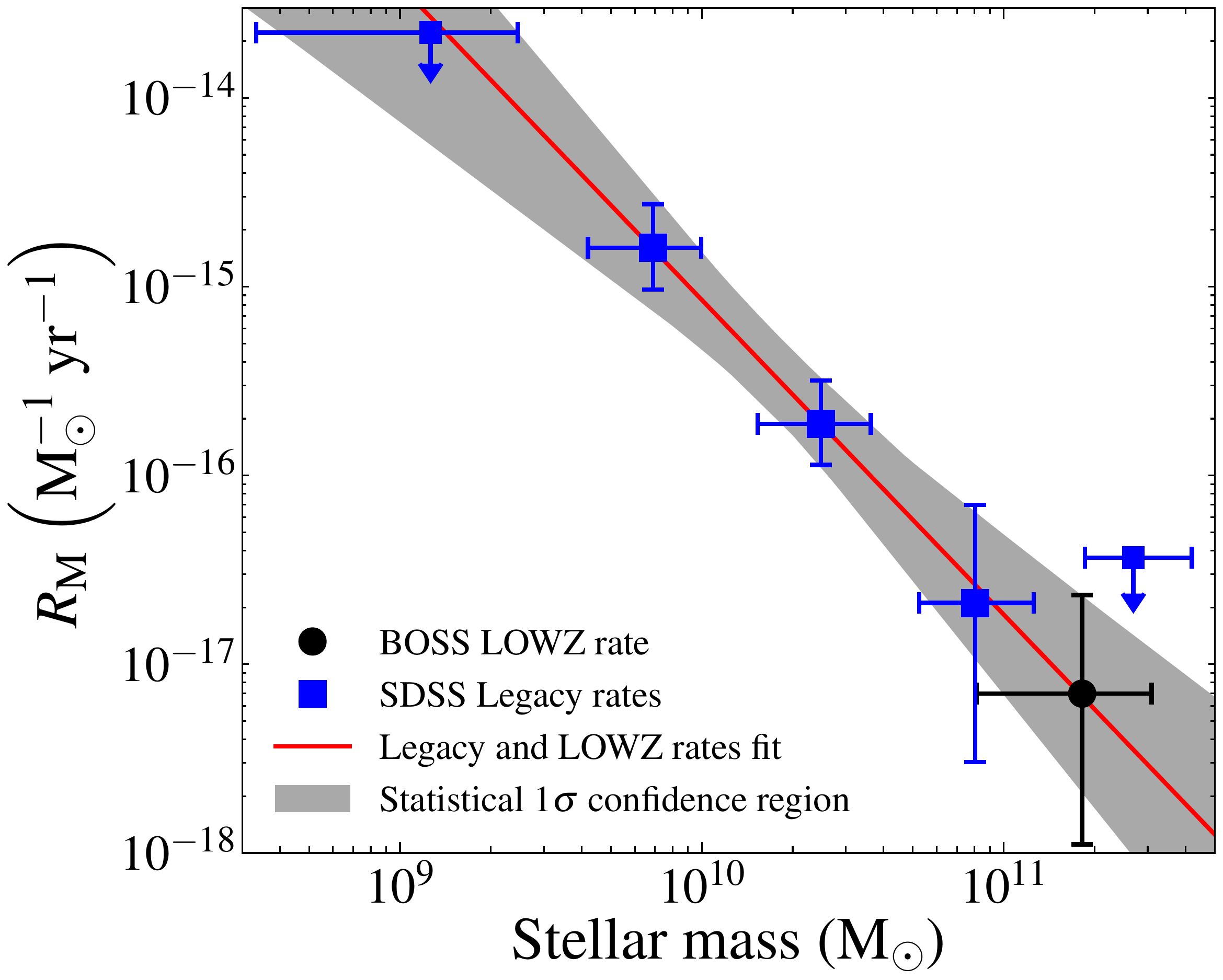}
	\caption{Mass-normalized vECLE rates as a function of galaxy stellar mass for SDSS Legacy (blue squares) and BOSS LOWZ (black circle).
	Vertical error bars show the statistical errors on the rates derived using the Monte Carlo simulations detailed above; and the horizontal error bars denote the range within each mass bin that 68 per cent of the galaxies fall.
	The points marked with downward arrows are $2\upsigma$ upper bounds on the rates calculated using the upper Poisson error on zero detections.
	The solid red line shows the power law fit to the Legacy and LOWZ rates and the shaded area is the $1\upsigma$ confidence region.}
	\label{fig:mass_rate_mass_relation}
\end{figure}

Using the same steps as for the galaxy-normalized rate vs. mass relation, we extend the mass-normalized rate vs. mass relation from \citeme\ to higher masses (\figref{fig:mass_rate_mass_relation}).
The stellar masses and mass-normalized rates of the SDSS Legacy bins and the LOWZ rate are shown in \tabref{tab:rate_mass_relations}.

A power law fit of the form of \equref{eq:rate_mass_relation} produces values of $a=-1.7~\pm0.3$ and $b=1.6^{+3.4}_{-3.5}$, with a reduced $\upchi^2=0.01$.
The resultant relation is consistent with the one measured by \citeme.

\begin{figure*}
	\centering
	\includegraphics[width=\textwidth]{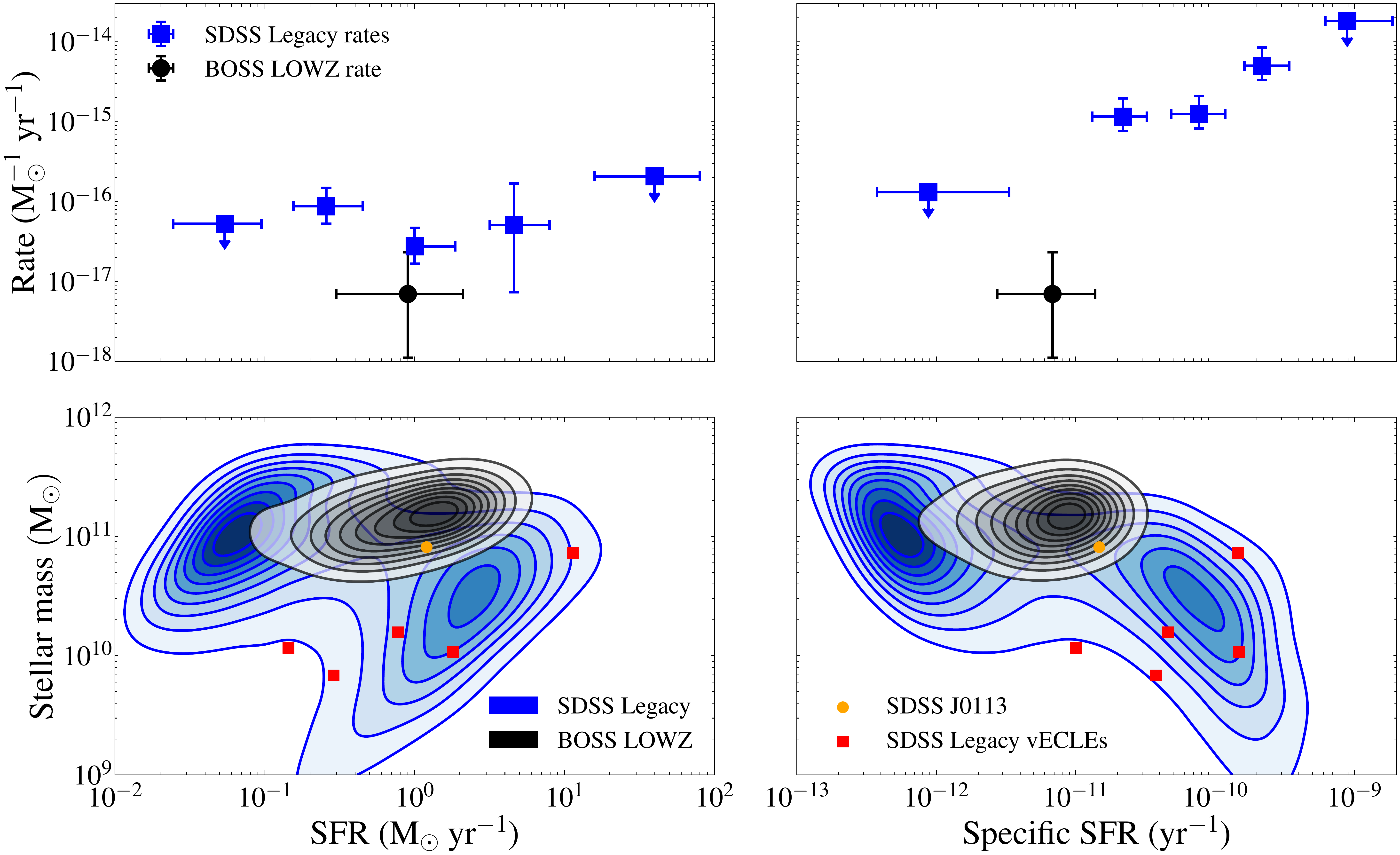}
	\caption{\textit{Top}: Mass-normalized vECLE rates as a function of galaxy SFR (left) and sSFR (right) for SDSS Legacy (blue squares) and BOSS LOWZ (black circle).
	Vertical error bars show the statistical errors on the rates, while the horizontal error bars denote the SFR and sSFR ranges comprising 68 per cent of the galaxies in each bin.
	The points marked with downward arrows are $2\upsigma$ upper limits on the rates calculated using the upper Poisson error on zero detections.
	\textit{Bottom}: Contours showing the density of galaxies in the SDSS Legacy (blue) and BOSS LOWZ (black) samples in decrements of 10 per cent.
	The vECLEs detected in SDSS Legacy by \citet{Wang2012} and \citeme\ are shown as red squares, and the vECLE detected in BOSS LOWZ, SDSS J0113, is shown as an orange circle.}
	\label{fig:rate_sfr_relation}
\end{figure*}

Following \citet{Graur2015}, we also investigate the dependence of the mass-normalized vECLE rate on the host galaxies' SFR and specific SFR (sSFR).
The sSFR is calculated by dividing the SFRs from the Portsmouth group pipeline by the corresponding masses.
We compare these rates to the mass-normalized vECLE rates from the SDSS Legacy survey (\citeme).
The masses and SFRs of the Legacy sample were derived by the MPA-JHU pipeline \citep{Kauffmann2003,Brinchmann2004,Tremonti2004}.
The Legacy galaxies are binned according to their SFRs, with three bins that contain one or two of the five vECLEs detected in that sample and two bins that contain no vECLEs, for which we calculate upper limits.
The rates for each of these bins are calculated as for the full sample, with the bins with no vECLEs calculated as $2\upsigma$ upper bounds on the rates using the Poisson error on 0 detections.
These rates are presented in Tables \ref{tab:rate_sfr_relations} and \ref{tab:rate_ssfr_relations} and \figref{fig:rate_sfr_relation}, along with the distributions of the SDSS Legacy and LOWZ samples.

\begin{table}
	\centering
	\caption{vECLE mass-normalized rates used in investigating the rate-SFR relation.}
	\begin{tabular}{cc}
		\hline
		SFR & Mass-normalized rate \\
		$\left(\msun~\mathrm{yr}^{-1}\right)$ & $\left(10^{-17}~\mathrm{M_\odot}^{-1}~\mathrm{yr}^{-1}\right)$ \\
		\hline
		\multicolumn{2}{c}{\textbf{SDSS Legacy}} \\
		$0.05~^{+0.04}_{-0.03}$ & $<5.3$ \\ [1mm]
		$0.25~^{+0.20}_{-0.10}$ & $8.7~^{+6.1}_{-3.4}$ \\ [1mm]
		$1.00~^{+0.86}_{-0.05}$ & $2.7~^{+2.0}_{-1.0}$ \\ [1mm]
		$4.6~^{+3.3}_{-1.4}$ & $5~^{+12}_{-4}$ \\ [1mm]
		$40~^{+40}_{-24}$ & $<21$ \\
		\multicolumn{2}{c}{\textbf{BOSS LOWZ}} \\
		$0.9~^{+1.2}_{-0.6}$ & $0.7~^{+1.6}_{-0.6}$ \\
		\hline
		\multicolumn{2}{l}{The rates marked with $<$ represent $2\upsigma$ upper limits.}
	\end{tabular}
	\label{tab:rate_sfr_relations}
\end{table}

\begin{table}
	\centering
	\caption{vECLE mass-normalized rates used in investigating the rate-sSFR relation.}
	\begin{tabular}{cc}
		\hline
		sSFR & Mass-normalized rate \\
		$\left(10^{-11}~\mathrm{yr}^{-1}\right)$ & $\left(10^{-17}~\mathrm{M_\odot}^{-1}~\mathrm{yr}^{-1}\right)$ \\
		\hline
		\multicolumn{2}{c}{\textbf{SDSS Legacy}} \\
		$0.09~^{+2.45}_{-0.05}$ & $<13$ \\ [1mm]
		$2.2~\pm0.1$ & $120~^{+80}_{-40}$ \\ [1mm]
		$8~^{+4}_{-3}$ & $120~^{+90}_{-40}$ \\ [1mm]
		$22~^{+12}_{-6}$ & $500~^{+350}_{-170}$ \\ [1mm]
		$90~^{+100}_{-30}$ & $<1900$ \\
		\multicolumn{2}{c}{\textbf{BOSS LOWZ}} \\
		$0.7~^{+0.7}_{-0.4}$ & $0.7~^{+1.6}_{-0.6}$ \\
		\hline
		\multicolumn{2}{l}{The rates marked with $<$ represent $2\upsigma$ upper limits.}
	\end{tabular}
	\label{tab:rate_ssfr_relations}
\end{table}

The mass-normalized rate shows no dependence on the SFR, with each bin having a very similar rate.
The LOWZ rate is consistent with the Legacy rate in the corresponding SFR bin.
\change{The mass-normalized vECLE rate appears to increase with sSFR.
We find that there may be a positive correlation for the four measurements that are not upper limits, though it is not statistically significant.
A likelihood ratio test prefers a 1\textsuperscript{st}-order polynomial over a 0\textsuperscript{th}-order polynomial at a $>2\upsigma$ confidence level.
In addition, we measure a Pearson r coefficient of 0.88 for these four measurements.
However, the p-value for this coefficient is 0.12, so we are unable to reject the null hypothesis that the rate is constant for the range of sSFRs.}

A correlation between SFR and CL emission has been presented by \citet{Molina2021} and \citet{Reefe2022}, who both found that CL emission is preferentially seen in galaxies with SFRs higher than predicted by the main sequence relationship for the galaxies' masses.
Given a key process that produces CL emission is accretion on to SMBHs, higher SFRs could enhance the feeding of these black holes, giving rise to CLs.
Combined with TDEs and vECLEs occurring preferentially in lower-mass galaxies (\citealt{Stone2016} and \citeme), this would result in vECLE rates being higher in galaxies with higher sSFRs.
\change{The gas required to reprocess high-energy emission to CLs is also more common in star-forming galaxies, which would further support this possible correlation.}
We also note that none of the six vECLEs from \citeme\ and this work are located in passive galaxies, again indicating that vECLEs are possibly linked to star formation.
\change{Given that the correlation we measure is not statistically significant, we require more measurements to fully test its validity.}

\subsection{Volumetric Rate}
\label{subsec:vol_rate}
Following \citet{Graur2013}, we calculate the volumetric vECLE rate by multiplying the mass-normalized rate by the total cosmic stellar density.
We determined this by integrating the galactic stellar mass function (GSMF) from \citet{Baldry2012} over the range of masses in the LOWZ sample, which is $10^{10}$ to $10^{12}~\mathrm{M}_\odot$ (\figref{fig:mass_hist}).
Though the \citet{Baldry2012} GSMF was only measured out to $z\sim0.06$, results from \citet{McLeod2021} and \citet{Hahn2024} have shown that the GSMF does not evolve significantly out to $z\sim0.5$.
Therefore, we are confident in using this GSMF for the LOWZ sample.
As can be seen in \figref{fig:mass_hist}, the LOWZ sample is biased towards higher mass galaxies, which in turn will have biased our vECLE sample.
We account for the relation between mass-normalized vECLE rate and galaxy stellar mass (\figref{fig:mass_rate_mass_relation}) by calculating the total cosmic mass density as the ratio of the integrated GSMF, $B\left(M\right)$, to the LOWZ sample galaxy-mass distribution, $D\left(M\right)$ (which is normalized such that $\int D(M)M\mathrm{d}M=1$), where both mass functions are weighted by the mass-normalized rate vs. mass relation, $R\left(M\right)$, determined in \secref{subsec:mass_rate}.
The resultant volumetric rate is
\begin{equation}
	R_\mathrm{V}=R_\mathrm{M}\frac{\int B(M)R(M)M\mathrm{d}M}{\int D(M)R(M)M\mathrm{d}M}.
	\label{eq:vol_rate}
\end{equation}

This method adds two additional sources of statistical uncertainty to our volumetric rate: the errors on the GSMF parameters determined by \citet{Baldry2012} and the uncertainty on the power law fit of the mass-normalized rate vs. mass relation in \secref{subsec:mass_rate}.
The resultant volumetric vECLE rate is \lowzvolrate.

To test the impact of each source of uncertainty, we repeat our Monte Carlo simulations, varying each source of uncertainty individually, and construct an error budget for the rates.
This is presented in \tabref{tab:uncertainties}.
The total uncertainty percentages are the linear sum of the total statistical and systematic uncertainties divided by the corresponding rate value.

When calculating these uncertainties, we did not consider the properties of the ISM which produces the 
CLs.
Variations in properties such as density, clumpiness, and ionization balance would increase the uncertainties presented here.

\begin{table}
	\centering
    \caption{ECLE rate uncertainty percentages.}
	\begin{tabular}{lc}
		\hline
		Uncertainty 				& Percentage of rate \\
		\hline
		\multicolumn{2}{c}{\textbf{Galaxy-normalized rate}} \\
		Total						& $+233/-84$ \\
		\multicolumn{2}{c}{\textit{Statistical}} \\
		Poisson						& $+229/-83$ \\
		AT~2017gge peak CL strength	& $\pm3$ \\
		AT~2017gge peak luminosity	& $+60/-15$ \\
		Range of power law indices	& $+23/-10$ \\
		\multicolumn{2}{c}{\textbf{Mass-normalized rate}} \\
		Total						& $+233/-84$ \\
		Galaxy stellar masses 		& $\pm1$ \\
		\multicolumn{2}{c}{\textbf{Volumetric rate}} \\
		Total						& $+249/-85$ \\
		GSMF parameters 			& $+112/-18$ \\
		Rate mass trend fit			& $+25/-18$ \\
		\hline
	\end{tabular}
	\label{tab:uncertainties}
\end{table}

\subsection{Comparison to Previous Work}
\label{subsec:rate_comp}
\begin{figure*}
	\centering
	\includegraphics[width=\textwidth]{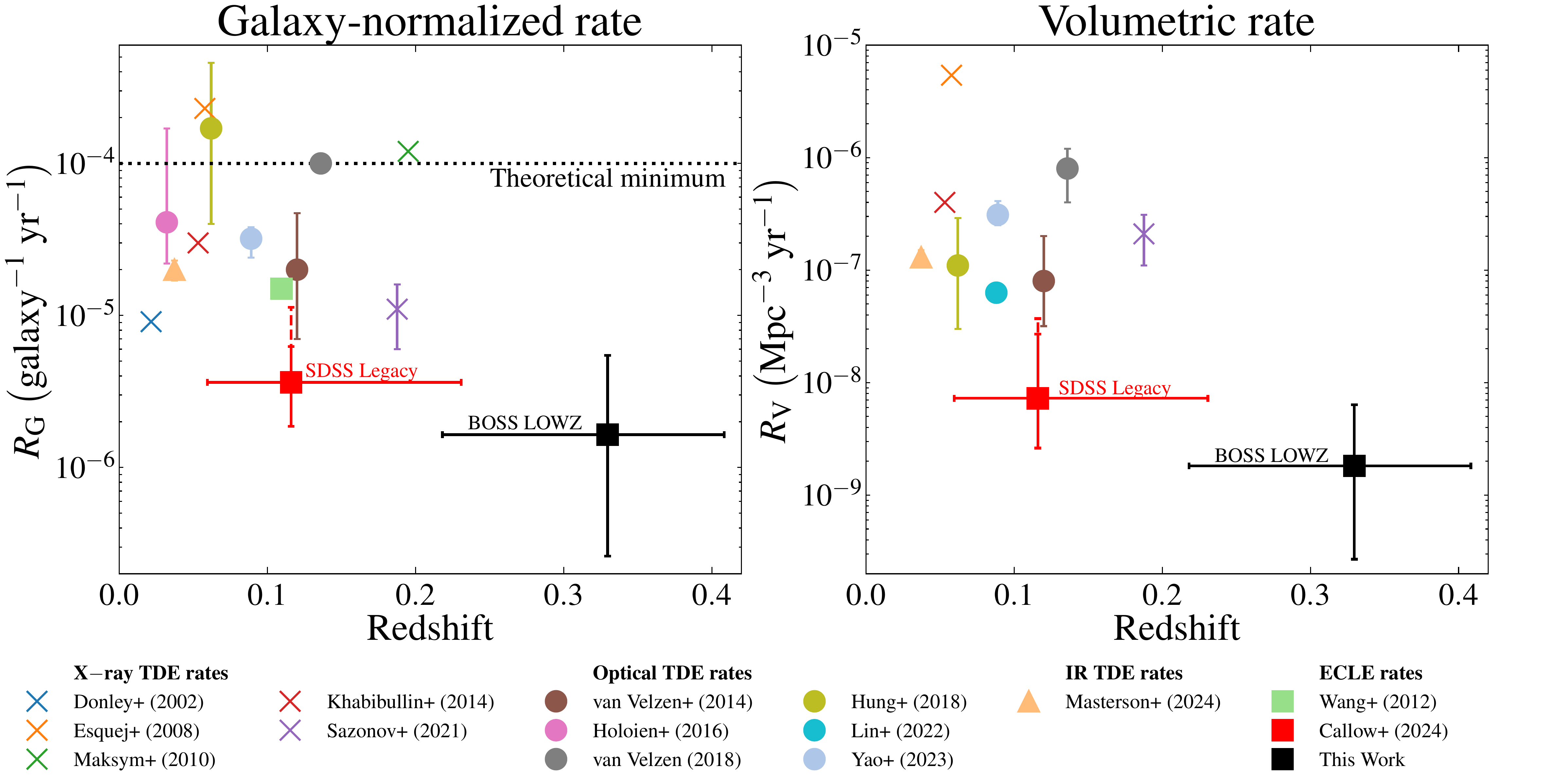}
	\caption{Comparisons of our galaxy-normalized (left) and volumetric (right) vECLE rates with TDE and ECLE rates from the literature.
	TDE rates derived from X-ray surveys are shown as crosses \citep{Donley2002,Esquej2008,Maksym2010,Khabibullin2014,Sazonov2021}, those from optical/UV surveys are shown as circles \citep{vanVelzen2014,Holoien2016,vanVelzen2018,Hung2018,Lin2022,Yao2023}, IR surveys are shown as triangles \citep{Masterson2024}, and ECLE rates are shown as squares \citep[\citeme, this work]{Wang2012}.
	Error bars are shown if available.
	The statistical errors of \citeme\ and this work are denoted by the solid error bars and the systematic errors by the dashed error bars.
	The horizontal error bars show the range of redshifts spanned by 68 per cent of the galaxies in BOSS LOWZ and the SDSS Legacy samples used here and by \citeme, respectively.
	The dotted horizontal line marks the theoretical minimum TDE rate calculated by \citet{Wang2004}.}
	\label{fig:rate_comp}
\end{figure*}

In \figref{fig:rate_comp}, we compare our galaxy-normalized and volumetric vECLE rates to TDE rates from the literature \citep{Donley2002,Esquej2008,Maksym2010,Khabibullin2014,vanVelzen2014,Holoien2016,Hung2018,vanVelzen2018,Sazonov2021,Lin2022,Yao2023,Masterson2024}.
We also include ECLE rates from \citet{Wang2012} and \citeme.

By searching for ECLEs in the LOWZ sample, we have probed the vECLE rate at a higher redshift range than \citeme\ and TDE observational rates.
We find that both the galaxy-normalized and volumetric LOWZ rates are consistent with the SDSS Legacy rates.
The LOWZ galaxy-normalized rate is also an order of magnitude lower than the ECLE rate from \citet{Wang2012}.
This rate estimate included two ECLEs that were later found to be non-variable, unlike the full rate calculations done by \citeme\ and this work.
Therefore, it is not surprising that our rate is lower than the one estimated by \citet{Wang2012}.

Our rates are formally lower than those measured by \citeme, though consistent within the measured uncertainties.
More precise measurements are required to ascertain whether the vECLE rate indeed drops with increasing redshift.
If that is the case, it would align with \mnraschange{theoretical work on the evolution of the TDE rate with redshift, which is expected to decrease with increasing redshift \citep{Kochanek2016}.}
Such an alignment between observed vECLE rates and theoretical TDE rates would lend further support to the suggestion that TDEs are the progenitors of vECLEs.

TDEs have only been well studied in the local universe, so there are no observational rate estimates past $z\sim0.2$.
Comparing our vECLE rates to the lower-redshift TDE rates, we find that our ECLE rates are one to two orders of magnitude lower than the \desichange{observed} TDE rates.
\desichange{The galaxy-normalized vECLE rate is also two orders of magnitude lower than the theoretical minimum TDE rate from \citet{Wang2004}, although, as mentioned in \secref{sec:intro}, \citet{Teboul2023} have shown that this minimum can be lowered by including strong scattering interactions.}
By assuming that all vECLEs are produced by TDEs, we can estimate the fraction of TDEs that produce strong variable CLs.
A comparison to the lowest galaxy-normalized and volumetric TDE rates (\citealt{Donley2002} and \citealt{Lin2022}, respectively) results in upper limits on this fraction of $20^{+40}_{-15}$ per cent from the galaxy-normalized rate and $3^{+7}_{-2}$ per cent from the volumetric rate.
As expected, these fractions are lower than those calculated by \citeme, but are consistent within the uncertainties.
However, these comparisons are made between our LOWZ vECLE rates and TDE rates, which are at a lower redshift.
If TDE rates decline at higher redshifts, as predicted, then these fractions would increase and appear more similar to those calculated by \citeme.

\section{Conclusions}
\label{sec:conclusions}
In this work, we investigated the vECLE rate at a higher redshift range than previously done by searching for ECLEs in the BOSS LOWZ sample and using follow-up spectra and MIR observations to determine their variability. We discovered two candidate ECLEs in the 341,110 BOSS LOWZ galaxies.

Using follow-up spectra from DESI and GMOS, along with \textit{WISE} MIR observations, we determined that only one of the candidate ECLEs (SDSS J0113) was variable.

Using this vECLE, we measured the galaxy-normalized vECLE rate in LOWZ to be \lowzgalrate. We also measured a mass-normalized rate of \lowzmassrate and converted it to a volumetric rate of \lowzvolrate.

The vECLE rate decreases with increasing galactic stellar mass, which is consistent with the theoretical calculations that predict a declining TDE rate with increasing SMBH mass.

By calculating the vECLE rate in LOWZ, we have probed a higher redshift than previous work. We found that the rates between SDSS Legacy and LOWZ are consistent with each other, but the LOWZ rate is formally lower. If upheld by future studies, a vECLE rate that declines with increasing redshift would align with theoretical work that predicts a similar declining TDE rate.

\change{We found no correlation between vECLE rate and SFR, but there is a $2\upsigma$ linear correlation with sSFR. If validated with future measurements, such a correlation would align with evidence that elevated SFR can enhance accretion on to SMBHs, the inverse scaling of vECLE rates with galaxy mass, and the necessity of gas for the productions of CLs.}

The LOWZ vECLE rates are one to two orders of magnitude lower than TDE rates at lower redshifts, which suggests that vECLEs are produced by $5\mathdash20$ per cent of TDEs. This fraction will be better constrained once TDE rates have been calculated for higher redshift samples, and large spectroscopic galaxy surveys, such as DESI, produce larger vECLE samples.

\section*{Acknowledgements}

\mnraschange{We thank the anonymous referee for helpful discussions and comments.}

This work was supported by the Science \& Technology Facilities Council [grants ST/S000550/1 and ST/W001225/1].

Funding for the Sloan Digital Sky Survey (SDSS) and SDSS-II has been provided by the Alfred P. Sloan Foundation, the Participating Institutions, the National Science Foundation, the U.S. Department of Energy, the National Aeronautics and Space Administration, the Japanese Monbukagakusho, and the Max Planck Society, and the Higher Education Funding Council for England. The SDSS Website is http://www.sdss.org/.

The SDSS is managed by the Astrophysical Research Consortium (ARC) for the Participating Institutions. The Participating Institutions are the American Museum of Natural History, Astrophysical Institute Potsdam, University of Basel, University of Cambridge, Case Western Reserve University, The University of Chicago, Drexel University, Fermilab, the Institute for Advanced Study, the Japan Participation Group, The Johns Hopkins University, the Joint Institute for Nuclear Astrophysics, the Kavli Institute for Particle Astrophysics and Cosmology, the Korean Scientist Group, the Chinese Academy of Sciences (LAMOST), Los Alamos National Laboratory, the Max-Planck-Institute for Astronomy (MPIA), the Max-Planck-Institute for Astrophysics (MPA), New Mexico State University, Ohio State University, University of Pittsburgh, University of Portsmouth, Princeton University, the United States Naval Observatory, and the University of Washington.

This material is based upon work supported by the U.S. Department of Energy (DOE), Office of Science, Office of High-Energy Physics, under Contract No. DE-AC02-05CH11231, and by the National Energy Research Scientific Computing Center, a DOE Office of Science User Facility under the same contract. Additional support for DESI was provided by the U.S. National Science Foundation (NSF), Division of Astronomical Sciences under Contract No. AST-0950945 to the NSF's National Optical-Infrared Astronomy Research Laboratory; the Science and Technology Facilities Council of the United Kingdom; the Gordon and Betty Moore Foundation; the Heising-Simons Foundation; the French Alternative Energies and Atomic Energy Commission (CEA); the National Council of Humanities, Science and Technology of Mexico (CONAHCYT); the Ministry of Science, Innovation and Universities of Spain (MICIU/AEI/10.13039/501100011033), and by the DESI Member Institutions: \url{https://www.desi.lbl.gov/collaborating-institutions}. Any opinions, findings, and conclusions or recommendations expressed in this material are those of the author(s) and do not necessarily reflect the views of the U. S. National Science Foundation, the U. S. Department of Energy, or any of the listed funding agencies.

The authors are honored to be permitted to conduct scientific research on Iolkam Du'ag (Kitt Peak), a mountain with particular significance to the Tohono O'odham Nation.

This research has made use of NASA's Astrophysics Data System Bibliographic Services and the NASA/IPAC Infrared Science Archive, which is funded by the National Aeronautics and Space Administration (NASA) and operated by the California Institute of Technology. This publication also makes use of data products from NEOWISE, which is a project of the Jet Propulsion Laboratory/California Institute of Technology, funded by the Planetary Science Division of NASA. The CRTS survey is supported by the U.S. National Science Foundation (NSF) under grants AST-0909182 and AST-1313422.

Based on observations obtained at the international Gemini Observatory, a program of NSF's NOIRLab, processed using DRAGONS (Data Reduction for Astronomy from Gemini Observatory North and South), which is managed by the Association of Universities for Research in Astronomy (AURA) under a cooperative agreement with the National Science Foundation on behalf of the Gemini Observatory partnership: the National Science Foundation (United States), National Research Council (Canada), Agencia Nacional de Investigaci\'{o}n y Desarrollo (Chile), Ministerio de Ciencia, Tecnolog\'{i}a e Innovaci\'{o}n (Argentina), Minist\'{e}rio da Ci\^{e}ncia, Tecnologia, Inova\c{c}\~{o}es e Comunica\c{c}\~{o}es (Brazil), and Korea Astronomy and Space Science Institute (Republic of Korea). This work was enabled by observations made from the Gemini North telescope, located within the Maunakea Science Reserve and adjacent to the summit of Maunakea. We are grateful for the privilege of observing the Universe from a place that is unique in both its astronomical quality and its cultural significance.

\section*{Data Availability}
The data underlying this article are available in the article and in its online supplementary material through Zenodo at \citet{Callow2024b}. The reduced data and derived measurements in this article will be shared on reasonable request to the corresponding author.

\bibliographystyle{mnras}
\bibliography{lowz_ecles}

\bsp
\label{lastpage}
\end{document}